\documentclass[aps,prd,nofootinbib,showpacs,superscriptaddress, preprint]{revtex4-1}
\pdfoutput=1
\usepackage[utf8]{inputenc}
\usepackage{amsmath,amssymb}
\usepackage{epsfig}
\usepackage{graphicx}
\usepackage[usenames,dvipsnames]{color}
\usepackage{float}
\usepackage{bbold}
\usepackage[colorlinks,citecolor=blue]{hyperref}
\usepackage{color}
\usepackage{subfigure}

\begin{document}
\title{Fermion Dark Matter with $N_2$ Leptogenesis \\ in Minimal Scotogenic Model}

\author{Devabrat Mahanta}
\email{devab176121007@iitg.ac.in}
\affiliation{Department of Physics, Indian Institute of Technology Guwahati, Assam 781039, India}

\author{Debasish Borah}
\email{dborah@iitg.ac.in}
\affiliation{Department of Physics, Indian Institute of Technology Guwahati, Assam 781039, India}

\begin{abstract}
We study the possibility of singlet fermion dark matter (DM) and successful leptogenesis in minimal scotogenic model which also provides a common origin of DM and light neutrino masses. In this scenario, where the standard model (SM) is extended by three gauge singlet fermions and one additional scalar doublet, all odd under an in-built $Z_2$ symmetry, the lightest singlet fermion which also happens to be the lightest $Z_2$ odd particle, can be either thermal or non-thermal DM candidate depending upon the strength of its couplings with SM leptons and the $Z_2$ odd scalar doublet. In both the scenarios, the $Z_2$ odd scalar doublet plays a non-trivial role either by assisting coannihilation with thermal DM or by providing a source for non-thermal DM via decay. The heavier $Z_2$ odd singlet fermion produces a net lepton asymmetry through its out-of-equilibrium decay into SM leptons and $Z_2$ odd scalar doublet. We show that the requirement of producing the observed baryon asymmetry pushes the scale of leptogenesis in case of normal ordering of light neutrino masses to several order of magnitudes above TeV scale. In case of inverted ordering however, it is possible to have successful $N_2$ leptogenesis at a scale of few tens of TeV. Inclusion of lepton flavour effects lowers this scale of leptogenesis by around an order of magnitude in both the cases. Correct DM abundance can be realised either by thermal freeze-out or by freeze-in mechanism in different parts of the parameter space that can have interesting prospects for ongoing experiments.

\end{abstract}
\maketitle

\section{Introduction}
\label{sec1}
The fact that the present universe has a significant amount of mysterious, non-luminous, non-baryonic form of matter, also known as dark matter (DM), is supported by several observations \cite{Zwicky:1933gu, Rubin:1970zza, Clowe:2006eq, Aghanim:2018eyx, Tanabashi:2018oca}. In terms of density 
parameter $\Omega_{\rm DM}$ and $h = \text{Hubble Parameter}/(100 \;\text{km} ~\text{s}^{-1} 
\text{Mpc}^{-1})$, the present DM abundance is conventionally reported as \cite{Aghanim:2018eyx}:
$\Omega_{\text{DM}} h^2 = 0.120\pm 0.001$
at 68\% CL. Since none of the particles in the standard model (SM) can satisfy the requirements \cite{Taoso:2007qk} a typical DM candidate should satisfy, several beyond standard model (BSM) proposals have been put forward in the past few decades \cite{Feng:2010gw}. The most popular as well as the most widely studied framework among these proposals is the weakly interacting massive particle (WIMP) paradigm~\cite{Kolb:1990vq}. For a recent review of WIMP models, please see \cite{Arcadi:2017kky}. Recently, due to the non-observation of WIMP at different direct detection experiments like LUX \cite{Akerib:2016vxi}, PandaX-II \cite{Tan:2016zwf, Cui:2017nnn} and Xenon1T \cite{Aprile:2017iyp, Aprile:2018dbl}, another DM framework has gained attention where the interactions between DM and SM particles are much more weaker compared to WIMP \cite{Hall:2009bx}. Due to its interactions and the way it gets populated in the universe, such DM candidates are categorised as freeze-in (or feebly interacting) massive particle (FIMP) paradigm. The tiny couplings between DM and visible sector can be naturally realised either by higher dimensional operators \cite{Hall:2009bx, Elahi:2014fsa, McDonald:2015ljz} or through some UV complete renormalisable theories \cite{Biswas:2018aib}. 

Apart from the mystery of DM, another puzzling observation is the asymmetry in the visible sector: an excess of baryons over antibaryons. It is often quoted in terms of baryon to photon ratio \cite{Tanabashi:2018oca, Aghanim:2018eyx}
\begin{equation}
\eta_B = \frac{n_{B}-n_{\bar{B}}}{n_{\gamma}} = 6.1 \times 10^{-10}
\label{etaBobs}
\end{equation}
If the universe had started in a baryon symmetric manner without any need of specific initial conditions, there has to be some dynamical mechanism that has led to such an asymmetry in the present epoch. Such a dynamical mechanism has to satisfy certain conditions, known as Sakharov's conditions \cite{Sakharov:1967dj} in order to generate a net asymmetry. These conditions are (i) baryon number (B) violation, (ii) C and CP violation and (iii) departure from thermal equilibrium. However, all these conditions can not be satisfied simultaneously in required amounts within the SM alone, requiring BSM frameworks to account for the asymmetry. One possible way is to extend the SM by heavy particles whose out-of-equilibrium decay can lead to the generation of baryon asymmetry of the universe (BAU) \cite{Weinberg:1979bt, Kolb:1979qa}. One interesting way to implement this mechanism is popularly known as leptogenesis, proposed by Fukugita and Yanagida more than thirty years back \cite{Fukugita:1986hr}. For a review of leptogenesis, please see \cite{Davidson:2008bu}. In leptogenesis, an asymmetry is generated in the lepton sector first which later gets converted into baryon asymmetry through $(B+L)$-violating EW sphaleron transitions~\cite{Kuzmin:1985mm}. An interesting feature of this scenario is that the required lepton asymmetry can be generated through CP violating out-of-equilibrium decays of the same heavy fields that take part in the seesaw mechanism~\cite{Minkowski:1977sc, Mohapatra:1979ia, Yanagida:1979as, GellMann:1980vs, Glashow:1979nm, Schechter:1980gr} that explains the origin of tiny neutrino masses~\cite{Tanabashi:2018oca}, another observed phenomenon the SM fails to address.


Motivated by the above observed phenomena which the SM fails to explain, we consider a BSM framework where the SM is extended by three copies of $Z_2$ odd fermions singlet under SM gauge symmetries, and an additional scalar field similar to the Higgs doublet of the SM, but odd under the unbroken $Z_2$ symmetry. It is the minimal model belonging to the scotogenic framework proposed by Ma in 2006 \cite{Ma:2006km}. The salient feature of this framework is the way it connects the origin of light neutrino masses and DM. The unbroken $Z_2$ symmetry leads to a stable DM candidate while the $Z_2$ odd particles generate light neutrino masses at one loop level. Apart from this, the out-of-equilibrium decay of the heavy singlet fermions can generate the required lepton asymmetry, which can give rise to the observed BAU after electroweak sphaleron transitions. Recently the authors of \cite{Hugle:2018qbw, Borah:2018rca} studied the possibility of creating lepton asymmetry from the decay of lightest singlet fermion $(N_1)$ decay and found that the required asymmetry can be produced for $M_1 \sim 10$ TeV within a vanilla leptogenesis framework having hierarchical $Z_2$ odd singlet fermionic masses while satisfying the constraints from light neutrino masses\footnote{Note that this is a significant improvement over the usual Davidson-Ibarra bound $M_1 > 10^9$ GeV for vanilla leptogenesis in type I seesaw framework \cite{Davidson:2002qv}}. Earlier works reporting low scale leptogenesis with $M_1$ having mass around few tens of TeV can be found in \cite{Hambye:2009pw, Racker:2013lua} where the author considered a hierarchical spectrum of right handed neutrinos. TeV scale leptogenesis in this model with quasi-degenerate right handed neutrinos was also discussed in earlier works \cite{Kashiwase:2012xd, Kashiwase:2013uy}. A high scale leptogenesis version of this scenario was studied by the authors of \cite{Huang:2018vcr}. In order to allow the decay of the lightest singlet fermion, the neutral component of the $Z_2$ odd scalar doublet had to be the DM candidate in these scenarios. Here we consider another possibility where the lightest $Z_2$ odd singlet fermion is also the lightest $Z_2$ odd particle, and hence the DM candidate. In this scenario the heavier singlet fermion $N_2$ decay is primarily responsible for generating the required lepton asymmetry  \footnote{An alternate possibility with keV scale $N_1$ DM and leptogenesis from a combination of the Akhmedov-Rubakov-Smirnov (ARS) mechanism \cite{Akhmedov:1998qx} and scalar doublet decay \cite{Hambye:2016sby} was recently explored by the authors of \cite{Baumholzer:2018sfb}. Leptogenesis from annihilations of $Z_2$ odd particles in this model was studied by the authors of \cite{Borah:2018uci}.}. It should be noted that $N_2$ decay dominating leptogenesis in usual type I seesaw mechanism was discussed in several earlier works \cite{DiBari:2005st, Vives:2005ra, Blanchet:2008pw, He:2008cd, Antusch:2010ms, Blanchet:2011xq, DiBari:2014eqa, Zhang:2015qia, DiBari:2008mp, DiBari:2010ux, DiBari:2014eya, DiBari:2015oca, DiBari:2015svd}. In these scenarios, the right handed neutrino spectrum is hierarchical and $N_1$ is too light to generate a sizeable asymmetry (lighter than the Davidson-Ibarra upper bound \cite{Davidson:2002qv}). The next to lightest right handed neutrino $N_2$ can be heavy enough and can produce the correct asymmetry for some parameter space of the models. This vanilla $N_2$ leptogenesis scenario is however, different from ours as in our case $N_1$ is perfectly stable and can not decay. We find that our $N_2$ leptogenesis scenario is more constrained compared to the vanilla $N_1$ decay scenario in the scotogenic model \cite{Hugle:2018qbw, Borah:2018rca}, pushing the scale of leptogenesis towards slightly higher side. This is due to decrease in number of free parameters to govern the production of CP asymmetry and opening up of more wash-out terms which destroy the CP asymmetry produced. On the other hand, the DM phenomenology can be richer due to the possibility of either WIMP or FIMP scenario. Since DM is a gauge singlet, it is possible, in principle, to realise either WIMP or FIMP scenario depending upon the smallness of respective Yukawa couplings. Interestingly, we find that the nature of $N_1$ DM is closely connected to the strength of several washout terms which further dictate the scale of leptogenesis. We constrain the parameter space from the requirement of generating the lepton asymmetry from $N_2$ decay, correct relic abundance of $N_1$ DM either via freeze-out or freeze-in while at the same time satisfying the constraints from light neutrino mass and mixing. Although the scale of leptogenesis gets pushed up, there exists rich new physics close to TeV scale in terms of DM and $Z_2$ odd scalar doublet that can be tested at ongoing experiments.

The rest of the paper is organised as follows. In section \ref{sec2}, we describe the minimal scotogenic model, its particle spectrum and origin of light neutrino masses. In section \ref{sec3}, we summarise the basic ways of calculating dark matter abundance in freeze-out and freeze-in scenarios. In section \ref{sec4}, we discuss the basics of leptogenesis from $N_2$ decay followed by discussion of our results in section \ref{sec5}. We finally conclude in section \ref{sec6}.

\section{Scotogenic Model}
\label{sec2}
As pointed out earlier, we consider the minimal model belonging to the scotogenic framework in our study. It is an extension f the SM by three copies of SM-singlet fermions $N_i$ (with $i=1,2,3$) and one $SU(2)_L$-doublet scalar field $\eta$ (also called inert doublet), all being odd under an in-built and unbroken $Z_2$ symmetry, while the SM fields remain $Z_2$-even, i.e. under the $Z_2$-symmetry, we have  
\begin{align}
N_i \rightarrow -N_i, \quad  \eta  \rightarrow -\eta, \quad \Phi_{1} \rightarrow \Phi_{1}, \quad \Psi_{\rm SM} \to 
\Psi_{\rm SM} \, ,
\label{eq:Z2}
\end{align}
where $\Phi_1$ is the SM Higgs doublet and $\Psi_{\rm SM}$'s stand for the SM fermions. This $Z_2$ symmetry, though \textit{ad hoc}  in this minimal setup, could be realised naturally as a subgroup of a continuous gauge symmetry like $U(1)_{B-L}$ with non-minimal field content \cite{Dasgupta:2014hha,Das:2017ski}. 

The relevant Yukawa Lagrangian involving the lepton sector is
\begin{equation}\label{IRHYukawa}
{\cal L} \ \supset \ \frac{1}{2}(M_N)_{ij} N_iN_j + \left(Y_{ij} \, \bar{L}_i \tilde{\eta} N_j  + \text{h.c.} \right) \ . 
\end{equation}
The $Z_2$ symmetry also prevents the usual Dirac Yukawa term $\bar{L}\tilde{\Phi}_1 N$ involving the SM Higgs, and hence, the Dirac mass term in the seesaw mechanism. This eventually forbids the generation of light neutrino masses at tree level through the conventional type I seesaw mechanism \cite{Minkowski:1977sc, Mohapatra:1979ia, Yanagida:1979as, GellMann:1980vs, Glashow:1979nm, Schechter:1980gr}.

The scalar sector of the model is same as the inert Higgs doublet model (IHDM) \cite{Deshpande:1977rw}, a minimal extension of the SM by a $Z_2$ odd scalar doublet in order to accommodate a DM candidate~\cite{Ma:2006km, Dasgupta:2014hha, Cirelli:2005uq, Barbieri:2006dq, Ma:2006fn,  LopezHonorez:2006gr,  Hambye:2009pw, Dolle:2009fn, Honorez:2010re, LopezHonorez:2010tb, Gustafsson:2012aj, Goudelis:2013uca, Arhrib:2013ela, Diaz:2015pyv, Ahriche:2017iar}. 
The scalar potential of the model involving the SM Higgs doublet $\Phi_1$ and the inert doublet $\eta$ can be written as
\begin{align}
V(\Phi_1,\eta) & \ = \   \mu_1^2|\Phi_1|^2 +\mu_2^2|\eta|^2+\frac{\lambda_1}{2}|\Phi_1|^4+\frac{\lambda_2}{2}|\eta|^4+\lambda_3|\Phi_1|^2|\eta|^2 \nonumber \\
& \qquad +\lambda_4|\Phi_1^\dag \eta|^2 + \left[\frac{\lambda_5}{2}(\Phi_1^\dag \eta)^2 + \text{h.c.}\right] \, . \label {c}
\end{align}
As mentioned earlier, in order to ensure that none of the neutral components of the inert Higgs doublet $\eta$ acquire a nonzero VEV, $\mu_2^2 >0$ is assumed. 

After the EWSB, these two scalar doublets can be written in the following form in the unitary gauge:
\begin{equation}
\Phi_1 \ = \ \begin{pmatrix} 0 \\  \frac{ v +h }{\sqrt 2} \end{pmatrix} , \qquad \eta \ = \ \begin{pmatrix} H^\pm\\  \frac{H^0+iA^0}{\sqrt 2} \end{pmatrix} \, ,
\label{eq:idm}
\end{equation}
where $h$ is the SM-like Higgs boson, $H^0$ and $A^0$ are the CP-even and CP-odd scalars, and $H^\pm$ are the charged scalars from the inert doublet. The masses of the physical scalars at tree level can be written as
\begin{eqnarray}
m_h^2 & \ = \ & \lambda_1 v^2 ,\nonumber\\
m_{H^\pm}^2 & \ = \ & \mu_2^2 + \frac{1}{2}\lambda_3 v^2 , \nonumber\\
m_{H^0}^2 & \ = \ & \mu_2^2 + \frac{1}{2}(\lambda_3+\lambda_4+\lambda_5)v^2 \ = \ m^2_{H^\pm}+
\frac{1}{2}\left(\lambda_4+\lambda_5\right)v^2  , \nonumber\\
m_{A^0}^2 & \ = \ & \mu_2^2 + \frac{1}{2}(\lambda_3+\lambda_4-\lambda_5)v^2 \ = \ m^2_{H^\pm}+
\frac{1}{2}\left(\lambda_4-\lambda_5\right)v^2 \, .
\label{mass_relation}
\end{eqnarray}
Without any loss of generality, we consider $ \lambda_5 >0$ so that the CP-odd scalar is lighter than the CP-even one. Since lightest component of inert doublet is not the DM candidate in our scenario, we can have any mass ordering among its components. This will not change the analysis we are going to do in upcoming sections, however these possibilities can be distinguished by their signatures at collider experiments like the large hadron collider (LHC).

Light neutrino masses which arise at one loop level can be evaluated as ~\cite{Ma:2006km, Merle:2015ica}
\begin{align}
(M_{\nu})_{ij} \ & = \ \sum_k \frac{Y_{ik}Y_{jk} M_{k}}{32 \pi^2} \left ( \frac{m^2_{H^0}}{m^2_{H^0}-M^2_k} \: \text{ln} \frac{m^2_{H^0}}{M^2_k}-\frac{m^2_{A^0}}{m^2_{A^0}-M^2_k}\: \text{ln} \frac{m^2_{A^0}}{M^2_k} \right) \nonumber \\ 
& \ \equiv  \ \sum_k \frac{Y_{ik}Y_{jk} M_{k}}{32 \pi^2} \left[L_k(m^2_{H^0})-L_k(m^2_{A^0})\right] \, ,
\label{numass1}
\end{align}
where 
$M_k$ is the mass eigenvalue of the mass eigenstate $N_k$ in the internal line and the indices $i, j = 1,2,3$ run over the three neutrino generations as well as three copies of $N_i$. The function $L_k(m^2)$ is defined as 
\begin{align}
L_k(m^2) \ = \ \frac{m^2}{m^2-M^2_k} \: \text{ln} \frac{m^2}{M^2_k} \, .
\label{eq:Lk}
\end{align}
From the expressions for physical scalar masses given in equations \eqref{mass_relation}, we can write $m^2_{H^0}-m^2_{A^0}=\lambda_5 v^2$. Therefore, in the limit $\lambda_5 \to 0$, the neutral components of inert doublet $\eta$ become mass degenerate. Also, a vanishing $\lambda_5$ implies vanishing light neutrino masses which is expected as the $\lambda_5$-term in the scalar potential~\eqref{c} breaks lepton number by two units, when considered together with the SM-singlet fermions Lagrangian~\eqref{IRHYukawa}. Since setting $\lambda_5 \to 0$ allows us to recover the lepton number global symmetry, the smallness of $\lambda_5$ is technically natural in the 't Hooft sense~\cite{tHooft:1979rat}. We will see later that such small $\lambda_5$ is indeed required for certain scenarios in order to achieve the desired phenomenology.

As we will see in the upcoming sections, the requirement of correct DM phenomenology for $N_1$ DM significantly constrain the Yukawa couplings. In particular, the requirement of FIMP DM tightly constrains the Yukawa couplings involving $N_1$ very small $ \leq 10^{-8}$ while for WIMP DM the same Yukawa couplings should be of order one $\mathcal{O}(1)$. Accordingly, the parameter $\lambda_5$ has to be tuned in order to generate the correct light neutrino masses. It is important to ensure that the choice of Yukawa couplings as well as other parameters involved in light neutrino mass are consistent with the cosmological upper bound on the sum of neutrino masses, $\sum_i m_{i}\leq 0.11$ eV~\cite{Aghanim:2018eyx}, as well as the neutrino oscillation data~\cite{deSalas:2017kay, Esteban:2018azc}. In order to incorporate these constraints on model parameters, it is often useful to rewrite the neutrino mass formula given in equation \eqref{numass1} in a form resembling the type-I seesaw formula: 
\begin{align}
M_\nu \ = \ Y {\Lambda}^{-1} Y^T \, ,
\label{eq:nu2}
\end{align}
where we have introduced the diagonal matrix $\Lambda$ with elements
\begin{align}
 \Lambda_i \ & = \ \frac{2\pi^2}{\lambda_5}\zeta_i\frac{2M_i}{v^2} \, , \\
\textrm {and}\quad \zeta_i & \ = \  \left(\frac{M_{i}^2}{8(m_{H^0}^2-m_{A^0}^2)}\left[L_i(m_{H^0}^2)-L_i(m_{A^0}^2) \right]\right)^{-1} \, . \label{eq:zeta}
\end{align}
The light neutrino mass matrix~\eqref{eq:nu2} which is complex symmetric, can be diagonalised by the usual Pontecorvo-Maki-Nakagawa-Sakata (PMNS) mixing matrix $U$ \footnote{Usually, the leptonic mixing matrix is given in terms of the charged lepton diagonalising matrix $(U_l)$ and light neutrino diagonalising matrix $U_{\nu}$ as $U = U^{\dagger}_l U_{\nu}$. In the simple case where the charged lepton mass matrix is diagonal which is true in our model, we can have $U_l = \mathbb{1}$. Therefore we can write $U = U_{\nu}$.}, written in terms of neutrino oscillation data (up to the Majorana phases) as
\begin{equation}
U=\left(\begin{array}{ccc}
c_{12}c_{13}& s_{12}c_{13}& s_{13}e^{-i\delta}\\
-s_{12}c_{23}-c_{12}s_{23}s_{13}e^{i\delta}& c_{12}c_{23}-s_{12}s_{23}s_{13}e^{i\delta} & s_{23}c_{13} \\
s_{12}s_{23}-c_{12}c_{23}s_{13}e^{i\delta} & -c_{12}s_{23}-s_{12}c_{23}s_{13}e^{i\delta}& c_{23}c_{13}
\end{array}\right) U_{\text{Maj}}
\label{PMNS}
\end{equation}
where $c_{ij} = \cos{\theta_{ij}}, \; s_{ij} = \sin{\theta_{ij}}$ and $\delta$ is the leptonic Dirac CP phase. The diagonal matrix $U_{\text{Maj}}=\text{diag}(1, e^{i\alpha}, e^{i(\zeta+\delta)})$ contains the undetermined Majorana CP phases $\alpha, \zeta$. The diagonal light neutrino mass matrix is therefore,
\begin{align}
D_\nu \ = \ U^\dag M_\nu U^* \ = \ \textrm{diag}(m_1,m_2,m_3) \, .
\end{align}   
where the light neutrino masses can follow either normal ordering (NO) or inverted ordering (IO). Since the inputs from neutrino data are only in terms of the mass squared differences and mixing angles, it would be  
useful for our purpose to express the Yukawa couplings in terms of light neutrino parameters. This is possible through the 
Casas-Ibarra (CI) parametrisation \cite{Casas:2001sr} extended to radiative seesaw model \cite{Toma:2013zsa} which 
allows us to write the Yukawa coupling matrix satisfying the neutrino data as
\begin{align}
Y \ = \ U D_\nu^{1/2} R^{\dagger} \Lambda^{1/2} \, ,
\label{eq:Yuk}
\end{align}
where $R$ is an arbitrary complex orthogonal matrix satisfying $RR^{T}=\mathbb{1}$.

\section{Dark matter}
\label{sec3}
As pointed out earlier, the DM candidate in our model is the lightest $Z_2$ odd singlet fermion $N_1$. Being gauge singlet, the production mechanism of $N_1$ DM crucially depends upon its Yukawa couplings with the SM leptons and inert doublet $\eta$. Depending upon the size of these Yukawa couplings, one can either realise WIMP or FIMP type DM in our model.

For WIMP type DM which is produced thermally in the early universe, its thermal relic abundance can be obtained by solving the Boltzmann equation for the evolution of the DM number density $n_{\rm DM}$:
\begin{equation}
\frac{dn_{\rm DM}}{dt}+3Hn_{\rm DM} \ = \ -\langle \sigma v \rangle \left[n^2_{\rm DM} -(n^{\rm eq}_{\rm DM})^2\right],
\label{eq:BE}
\end{equation}
where $n^{\rm eq}_{\rm DM}$ is the equilibrium number density of DM and $ \langle \sigma v \rangle $ is the thermally averaged annihilation cross section, given by~\cite{Gondolo:1990dk}
\begin{equation}
\langle \sigma v \rangle \ = \ \frac{1}{8m_{\rm DM}^4T K^2_2\left(\frac{m_{\rm DM}}{T}\right)} \int\limits^{\infty}_{4m_{\rm DM}^2}\sigma (s-4m_{\rm DM}^2)\sqrt{s}\: K_1\left(\frac{\sqrt{s}}{T}\right) ds \, ,
\label{eq:sigmav}
\end{equation}
where $K_i(x)$'s are modified Bessel functions of order $i$. In the presence of coannihilation, one follows the recipe given by~\cite{Griest:1990kh} to calculate the relic abundance. For a recent study on such coannihilation effects on fermion DM in this model, please see \cite{Borah:2018smz}. As we will show later, the requirement of successful $N_2$ leptogenesis pushes the masses of $N_{2,3}$ to higher values, making their coannihilations with $N_1$ highly inefficient. However, the mass of $\eta$ can remain very close to that of $N_1$ enhancing the coannihilation effects.

On the other hand, if the Yukawa couplings of $N_1$ with SM leptons are very small, the FIMP possibility will arise. In such a case, as mentioned earlier, $N_1$ never reaches thermal equilibrium with the standard bath and has to be generated from decay or scattering of particles in the thermal bath. If the same couplings are involved in both scattering and decay, then decay contributions dominate \cite{Hall:2009bx}. In our model, the most dominant decay producing $N_1$ is the two body decay of $\eta \rightarrow l N_1$ given by
\begin{equation}
\Gamma_{\eta\rightarrow N_{1} l}\cong\dfrac{m_{\eta}Y^{2}}{8\pi}\left (1-\dfrac{M_{1}^{2}}{m_{\eta}^{2}}\right )^{2}
\label{decayeta1}
\end{equation}
where $Y$ is the effective Yukawa coupling (up to the flavour indices), $M_{1}$ is the mass of FIMP type DM particle $N_{1}$ and $m_{\eta}$ is the mass of the mother particle. By virtue of its gauge interactions, $\eta$ can be thermally produced in the early universe. Therefore, the coupled Boltzmann equations for comoving number densities of $N_1$ and $\eta$ can be written as 
\begin{align}
\dfrac{dY_{\eta}}{dz}&=-\dfrac{4\pi^{2}}{45}\dfrac{M_{\rm Pl}m_{\eta}}{1.66}\dfrac{\sqrt{g_{*}(z)}}{z^{2}}\bigg [\displaystyle\sum_{p \equiv \rm SM \;particles} \langle \sigma  v \rangle_{\eta\eta\rightarrow pp}(Y_{\eta}^{2}-(Y_{\eta}^{\rm eq})^{2} \bigg )]\\&-\dfrac{M_{\rm Pl}}{1.66}\dfrac{z}{m_{\eta}^{2}}\dfrac{\sqrt{g_{*}(z)}}{g_{s}(z)}\Gamma_{\eta\rightarrow N_{1} l}Y_{\eta}
\label{eq:29}
\end{align}
\begin{align}
\dfrac{dY_{N_{1}}}{dz}&=\dfrac{M_{\rm Pl}}{1.66}\dfrac{z}{m_{\eta}^{2}}\dfrac{\sqrt{g_{*}(z)}}{g_{s}(z)}\Gamma_{\eta\rightarrow N_{1} l}Y_{\eta}
\label{eq:29b}
\end{align}
where $z=\dfrac{m_{\eta}}{T}$ is a dimensionless variable and $M_{\rm Pl}$ is the Planck mass. $g_{s}(z)$ is the number of effective relativistic degrees of freedom associated with the entropy density of the universe at some $z$, and the $g_{*}(z)$ is defined by
\begin{eqnarray}
\sqrt{g_{\star}(z)} = \dfrac{g_{\rm s}(z)}
{\sqrt{g_{\rho}(z)}}\,\left(1 -\dfrac{1}{3}
\dfrac{{\rm d}\,{\rm ln}\,g_{\rm s}(z)}{{\rm d} \,{\rm ln} z}\right)\,. 
\end{eqnarray}
Here, $g_{\rho}(x)$ denotes the effective number of degrees
of freedom related to the energy density of the universe at
$z$. 

\section{Leptogenesis}
\label{sec4}
As mentioned earlier, a net lepton asymmetry can be generated in this model via out-of-equilibrium decay of the $N_i$~\cite{Ma:2006fn, Kashiwase:2012xd, Kashiwase:2013uy, Racker:2013lua, Clarke:2015hta, Hugle:2018qbw, Borah:2018rca}. Similar to the Davidson-Ibarra bound in type I seesaw leptogenesis mentioned earlier, here also one can derive a comparable lower bound with only two $Z_2$ odd singlet fermions in the strong washout regime. With three singlet fermions in the scotogenic model, this bound can be lowered down to around 10 TeV \cite{Hugle:2018qbw, Borah:2018rca} without any need of resonance enhancement \cite{Pilaftsis:2003gt, Dev:2017wwc}. Since we consider the leptogenesis to be generated from $N_2$ decay effectively, by considering $N_1$ to be the lightest $Z_2$ odd particle which can not decay, our scenario is more constrained compared to the ones discussed in \cite{Hugle:2018qbw, Borah:2018rca}. Although $N_3$ decay can also generate lepton asymmetry, in principle, we consider the asymmetry generated by $N_3$ decay or any pre-existing asymmetry to be negligible due to strong washout effects mediated either by $N_{2}$ or $N_{3}$ themselves. We also neglect $\Delta L=1$ scattering processes and flavour effects.

The CP asymmetry parameter is defined as
\begin{equation}
\epsilon_{i} =\frac{\sum_{\alpha}\Gamma(N_{i}\rightarrow l_{\alpha}\eta)-\Gamma(N_{i}\rightarrow\bar{l_{\alpha}}\bar{\eta})}{\sum_{\alpha}\Gamma(N_{i}\rightarrow l_{\alpha}\eta)+\Gamma(N_{i}\rightarrow\bar{l_{\alpha}}\bar{\eta})}.
\label{epsilon1}
\end{equation} 
\\
The CP asymmetry parameter for $N_i \rightarrow l_{\alpha} \eta, \bar{l_{\alpha}}\bar{\eta}$ is given by 
\begin{equation}
\epsilon_{i \alpha} = \frac{1}{8 \pi (Y^{\dagger}Y)_{ii}} \sum_{j\neq i} \bigg [ f \left( \frac{M^2_j}{M^2_i}, \frac{m^2_{\eta}}{M^2_i} \right) {\rm Im} [ Y^*_{\alpha i} Y_{\alpha j} (Y^{\dagger} Y)_{ij}] - \frac{M^2_i}{M^2_j-M^2_i} \left( 1-\frac{m^2_{\eta}}{M^2_i} \right)^2 {\rm Im}[Y^*_{\alpha i} Y_{\alpha j} H_{ij}] \bigg ]
\label{epsilonflav}
\end{equation}
where, the function $f(r_{ji},\eta_{i})$ is coming from the interference of the tree-level and one loop diagrams and has the form
\begin{equation}
f(r_{ji},\eta_{i})=\sqrt{r_{ji}}\left[1+\frac{(1-2\eta_{i}+r_{ji})}{(1-\eta_{i}^{2})^{2}}{\rm ln}(\frac{r_{ji}-\eta_{i}^{2}}{1-2\eta_{i}+r_{ji}})\right]
\end{equation}
with $r_{ji}=M_{j}^{2}/M_{i}^{2}$ and $\eta_{i}=m_{\eta}^{2}/M_{i}^{2}$. The self energy contribution $H_{ij}$ is given by 
\begin{equation}
H_{ij} = (Y^{\dagger} Y)_{ij} \frac{M_j}{M_i} + (Y^{\dagger} Y)^*_{ij}
\end{equation}
Now, the CP asymmetry parameter, neglecting the flavour effects (summing over final state flavours $\alpha$) is
\begin{equation}
\epsilon_{i}=\frac{1}{8\pi(Y^{\dagger}Y)_{ii}}\sum_{j\neq i}{\rm Im}[((Y^{\dagger}Y)_{ij})^{2}]\frac{1}{\sqrt{r_{ji}}}F(r_{ji},\eta_{i})
 \label{eq:14}
\end{equation}
\\
where the function $F(r_{ji},\eta)$ is defined as 
\begin{equation}
F(r_{ji},\eta_{i})=\sqrt{r_{ji}}\left[ f(r_{ji},\eta_{i})-\frac{\sqrt{r_{ji}}}{r_{ji}-1}(1-\eta_{i})^{2} \right].
\end{equation}
Let us define the decay parameter as
\begin{equation}
K_{N_2}=\dfrac{\Gamma_{2}}{H(z=1)}
\end{equation}
where $\Gamma_{2}$ is the $N_{2}$ decay width, $H$ is the Hubble parameter and $z=M_{2}/T$ with $T$ being the temperature of the thermal bath. Leptogenesis occurs far above the electroweak scale where the universe was radiation dominated. In this era the Hubble parameter can be expressed in terms of the temperature $T$ as follows
\begin{equation}
H=\sqrt{\dfrac{8\pi^{3} g_{*}}{90}}\dfrac{T^{2}}{M_{Pl}}=H(z=1)\dfrac{1}{z^{2}}
\end{equation}
where $g_{*}$ is the effective number of relativistic degrees of freedom and $M_{\rm Pl}\simeq 1.22\times10^{19}$ GeV is the Planck mass. The decay width $\Gamma_{2}$ can be calculated as 
\begin{equation}
\Gamma_{2}=\dfrac{M_{2}}{8\pi}(Y^{\dagger} Y)_{22}(1-\eta_{2})^{2}
\end{equation}
The frequently appearing $Y^{\dagger}Y$ is calculated using Casas-Ibarra parametrisation and it is given as 
\begin{equation}
(Y^{\dagger}Y)_{ij}=\sqrt{\Lambda_{i}\Lambda_{j}}(RD_{\nu}R^{\dagger})_{ij}
\label{eq:20}
\end{equation}
$D_{\nu}={\rm diag}(m_{1},m_{2},m_{3})$ is the diagonal active neutrino mass matrix. One important point here is to note down that the important quantity $Y^{\dagger}Y$ for leptogenesis is independent of the lepton mixing PMNS matrix, whereas it is dependent on the complex angles of the CI parametrization. Thus the CP violating phases relevant for leptogenesis are independent of the CP violating phases in the PMNS matrix. The dependence of the CP asymmetry on $M_{i}$ and $\lambda_{5}$ is evident through $\Lambda_{i}$.

The basic equations to track the dynamics of leptogenesis are the Boltzmann equations given by~\cite{Buchmuller:2004nz}
\begin{eqnarray}
\frac{dn_{N_2}}{dz}& \ = \ &-D_2 (n_{N_2}-n_{N_2}^{\rm eq}) \, , \label{eq:bol1} \\
\frac{dn_{B-L}}{dz}& \ = \ &-\epsilon_2 D_2 (n_{N_2}-n_{N_2}^{\rm eq})-W^{\rm Total}n_{B-L} \, , \label{eq:bol2}
\end{eqnarray}
where $n_{N_2}^{\rm eq}=\frac{z^2}{2}K_2(z)$ is the equilibrium number density of $N_1$ (with $K_i(z)$ being the modified Bessel function of $i$-th kind). The quantity on the right hand side of the above equations
\begin{align}
D_2 \ \equiv \ \frac{\Gamma_{2}}{Hz} \ = \ K_{N_2} z \frac{K_1(z)}{K_2(z)}
\end{align}
 measures the total decay rate of $N_2$ with respect to the Hubble expansion rate, and similarly, $W^{\rm Total} \equiv \frac{\Gamma_{W}}{Hz}$ measures the total washout rate. The washout term is the sum of two contributions, i.e. $W^{\rm Total}=W_1+W_{\Delta L}$, where the washout due to the inverse decays $\ell \eta$, $\bar\ell \eta^* \rightarrow N_2$ is given by 
\begin{align}
W_1=W_{\rm ID} \ = \ \frac{1}{4}K_{N_2} z^3 K_1(z).
\end{align}
The other contribution to washout $ W_{\Delta L}$ originates from scatterings which violate lepton number by $\Delta L=1, 2$. The contribution from $\Delta L=2$ scatterings $\ell \eta \leftrightarrow \bar{\ell} \eta^*$, $\ell\ell\leftrightarrow \eta^* \eta^*$ is given by~\cite{Hugle:2018qbw} 
\begin{align}
W_{\Delta L=2} \ \simeq \ \frac{18\sqrt{10}\,M_{\rm Pl}}{\pi^4 g_\ell \sqrt{g_*}z^2 v^4}\left(\frac{2\pi^2}{\lambda_5}\right)^2 M_{2}\bar{m}_\zeta^2 \, ,
\label{eq:wash2}
\end{align}
where we have assumed $\eta_2\ll 1$ for simplicity, $g_\ell$ stands for the internal degrees of freedom for the SM leptons, and  $\bar{m}_\zeta$ is the effective neutrino mass parameter, defined as 
\begin{align}
\bar{m}_\zeta^2 \  \simeq \  4\zeta_1^2 m_{l}^2+\zeta_2 m_{h_1}^2+\zeta_3^2 m_{h_2}^2 \, ,
\end{align}
with $m_l, m_{h_1, h_2}$ are being the lightest and heavier neutrino mass eigenvalues, $\zeta_i$ defined in equation~\eqref{eq:zeta} and $L_i(m^2)$ defined in equation~\eqref{eq:Lk}. It should be noted that equation~\eqref{eq:wash2} is similar to the $\Delta L=2$ washout term in vanilla leptogenesis, except for the $\left(\frac{2\pi^2}{\lambda_5}\right)^2$ factor. In $N_2$ leptogenesis, which we discuss here, there can be other scattering diagrams which can significantly contribute to the washout processes. We include all of them in our work. One possible $\Delta L=2$ washout relevant at $T=M_2$ is from the processes $\ell \ell \longrightarrow N_{1}N_{1}$ which can be written as 
\begin{equation}
 W^{(1)}_{\Delta L=2}=\dfrac{zs}{H(z=1)}n_{\ell}^{\rm eq}r_{N_{2}}^{2} \langle \sigma v \rangle_{N_{1}N_{1}\longleftrightarrow \ell \ell}
\end{equation}
where $s$ is the entropy density in radiation dominated universe. The other two relevant $\Delta L=2$ washout contributions can be written as
\begin{equation}
W^{(2)}_{\Delta L=2}=\dfrac{z s}{H(z=1)}(n_{\eta}^{\rm eq} \langle \sigma v \rangle_{\eta l \longleftrightarrow \eta^{*} \bar{\ell}}+n_{N_{1}}^{\rm eq} \langle \sigma v \rangle_{N_{1} \ell \longleftrightarrow N_{1} \bar{\ell}})
\end{equation}
where $r_{j}=\dfrac{n_{j}^{\rm eq}}{n_{l}^{\rm eq}}$. The contribution from $\Delta L=1$ washout processes can be estimated as 
\begin{equation}
W_{\Delta L=1}=\dfrac{z s}{H(z=1)}n_{\ell}^{\rm eq}r_{N_{1}}r_{\eta}(\langle \sigma v \rangle_{\ell W^{\pm} \longleftrightarrow N_{1} \eta^{\pm}}+ \langle \sigma v \rangle_{\ell Z \longleftrightarrow N_{1} \eta^{0}})
\end{equation}

After obtaining the numerical solutions of the above Boltzmann equations \eqref{eq:bol1} and \eqref{eq:bol2}, we convert the final $B-L$ asymmetry $n_{B-L}^f$ just before electroweak sphaleron freeze-out into the observed baryon to photon ratio by the standard formula 
\begin{align}
\eta_B \ = \ \frac{3}{4}\frac{g_*^{0}}{g_*}a_{\rm sph}n_{B-L}^f \ \simeq \ 9.2\times 10^{-3}\: n_{B-L}^f \, ,
\label{eq:etaB}
\end{align}
where $a_{\rm sph}=\frac{8}{23}$ is the sphaleron conversion factor (taking into account two Higgs doublets). We take the effective relativistic degrees of freedom to be $g_*=110.75$, slightly higher than that of the SM at such temperatures as we are including the contribution of the inert doublet too. In the WIMP DM scenario it will be enhanced by approximately 1 as $N_1$ remains in thermal equilibrium. The heavier singlet fermions $N_{2,3}$ do not contribute as they have already decoupled from the bath by this epoch. In the above expression $g_*^0=\frac{43}{11}$ is the effective relativistic degrees of freedom at the recombination epoch.
\begin{figure}[H]
\centering
\includegraphics[scale=.6]{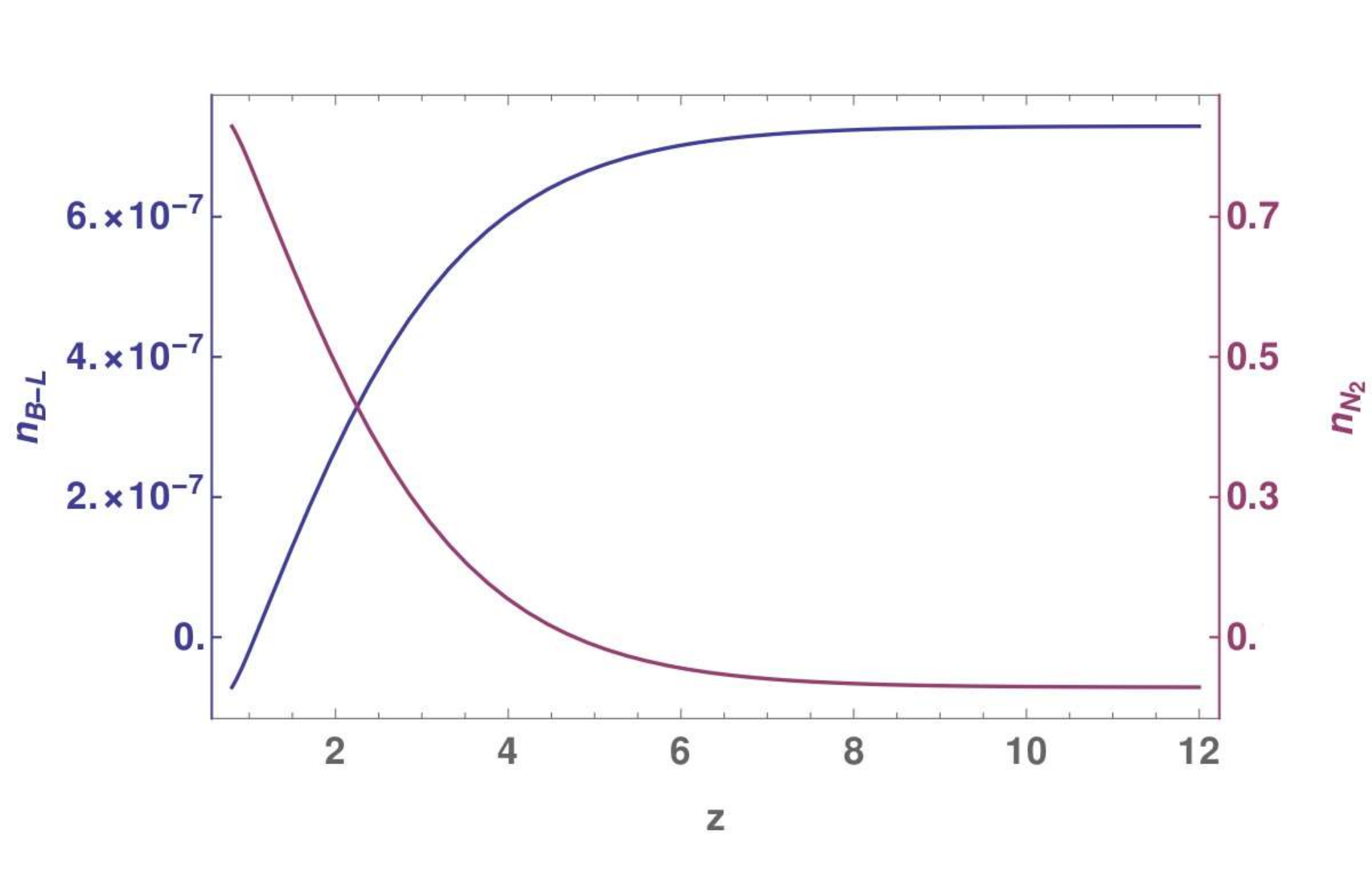}
\caption{Evolution $n_{N_2}, n_{B-L}$ (Comoving number densities of of $N_2$, $B-L$) with $z$ for normal ordering. The set of parameters used are $M_{2}=4 \times 10^{7}$ GeV, $M_{3}/M_{2}=10^{5}$, $M_{1}=201$ GeV, $m_{\eta}=201$ GeV, $\lambda_{5}=5.0, m_1 = 10^{-13}$ eV.}
\label{leptoNH1}
\end{figure}


\section{Results and discussion}
\label{sec5}  
\begin{figure}
\includegraphics[scale=.3]{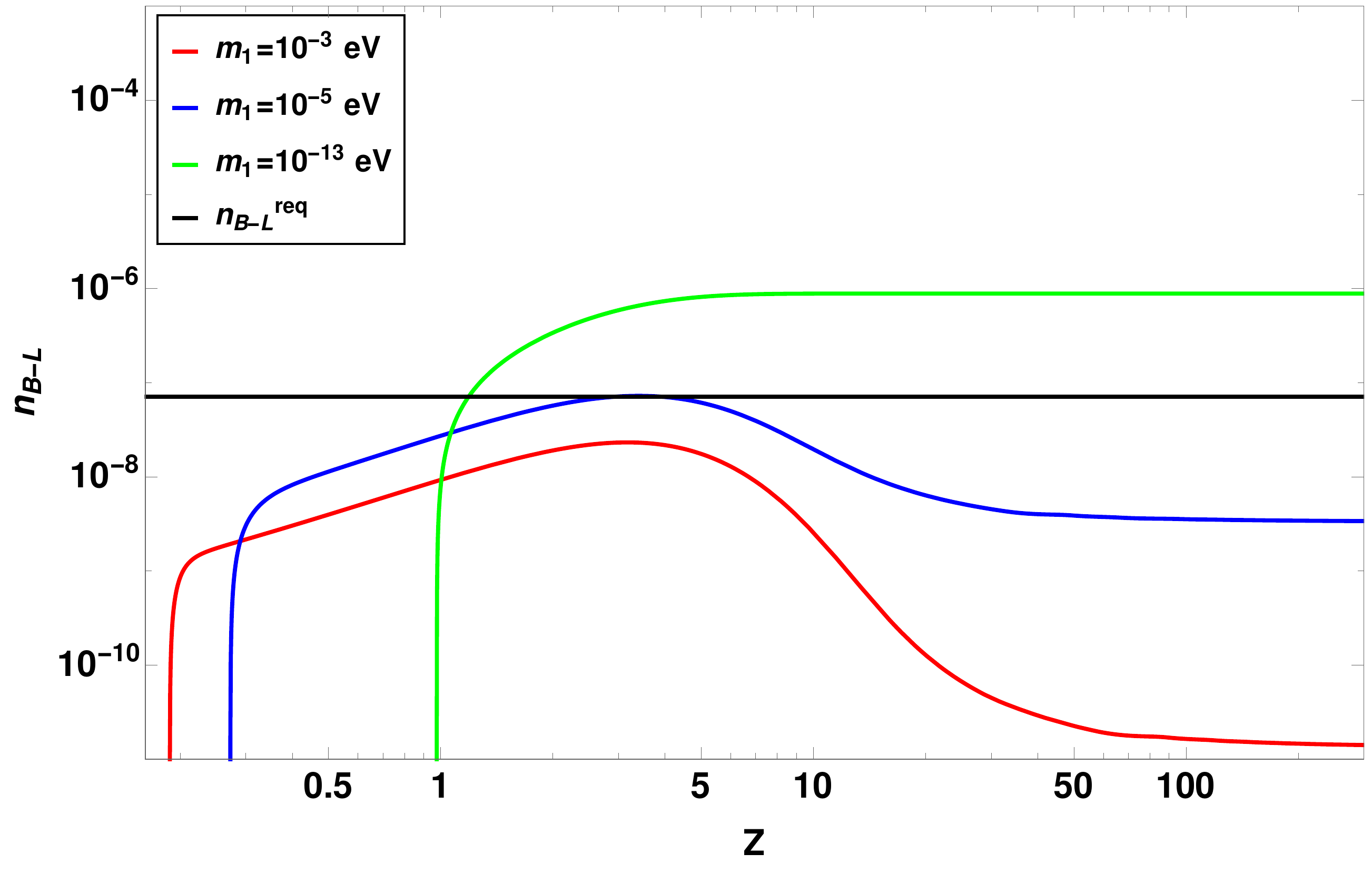}
\includegraphics[scale=.3]{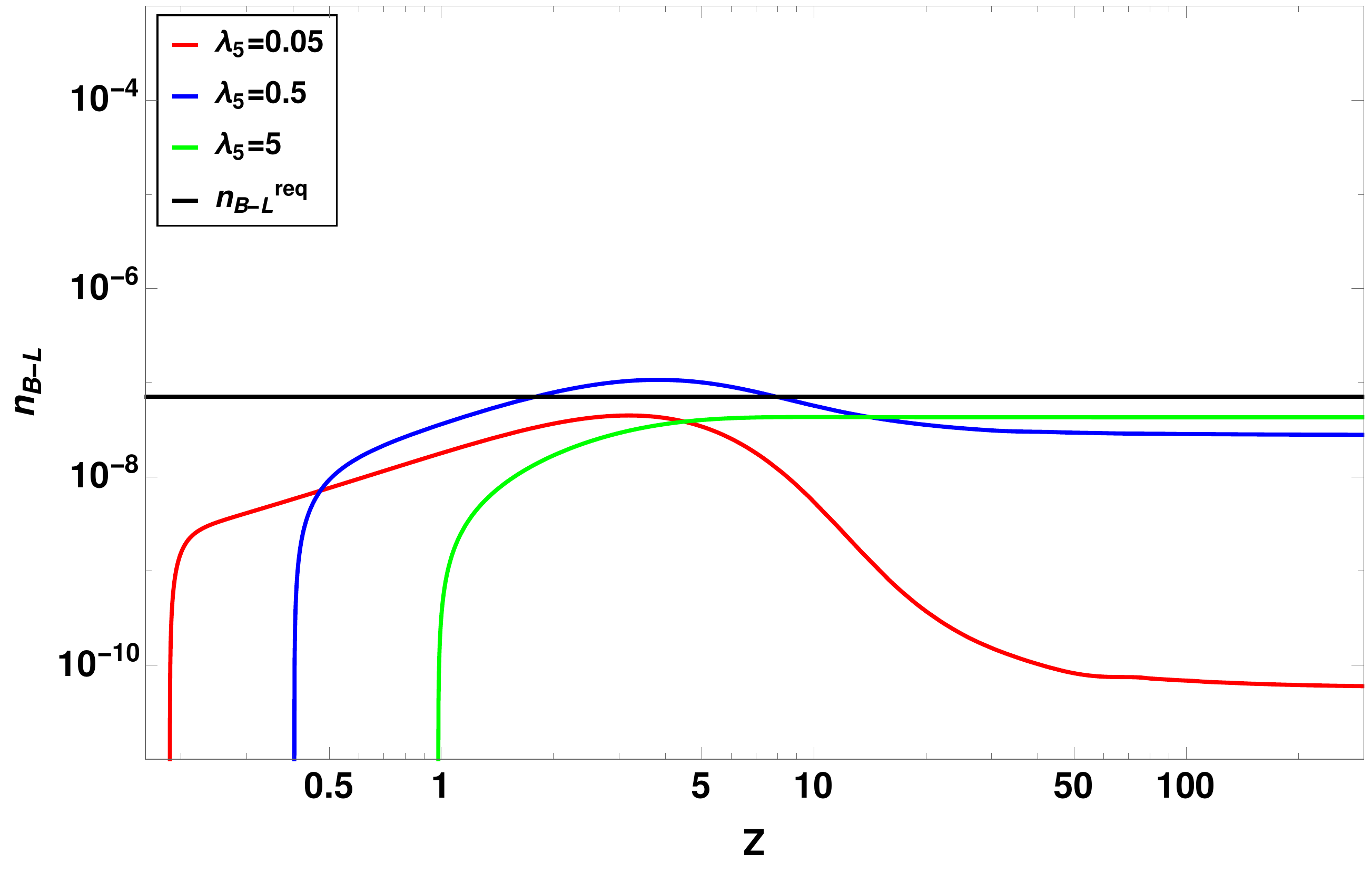} \\
\includegraphics[scale=.3]{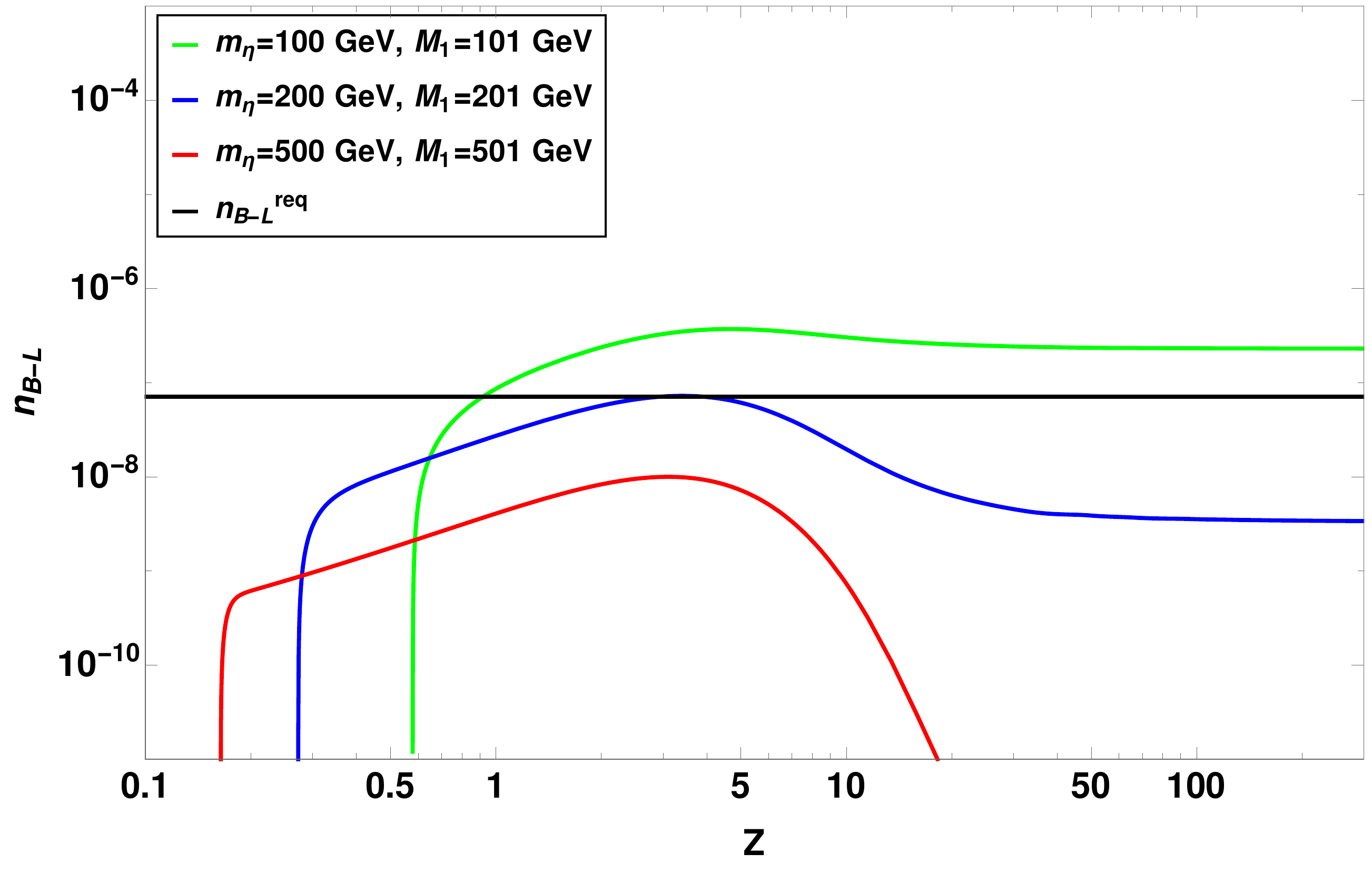}
\includegraphics[scale=.22]{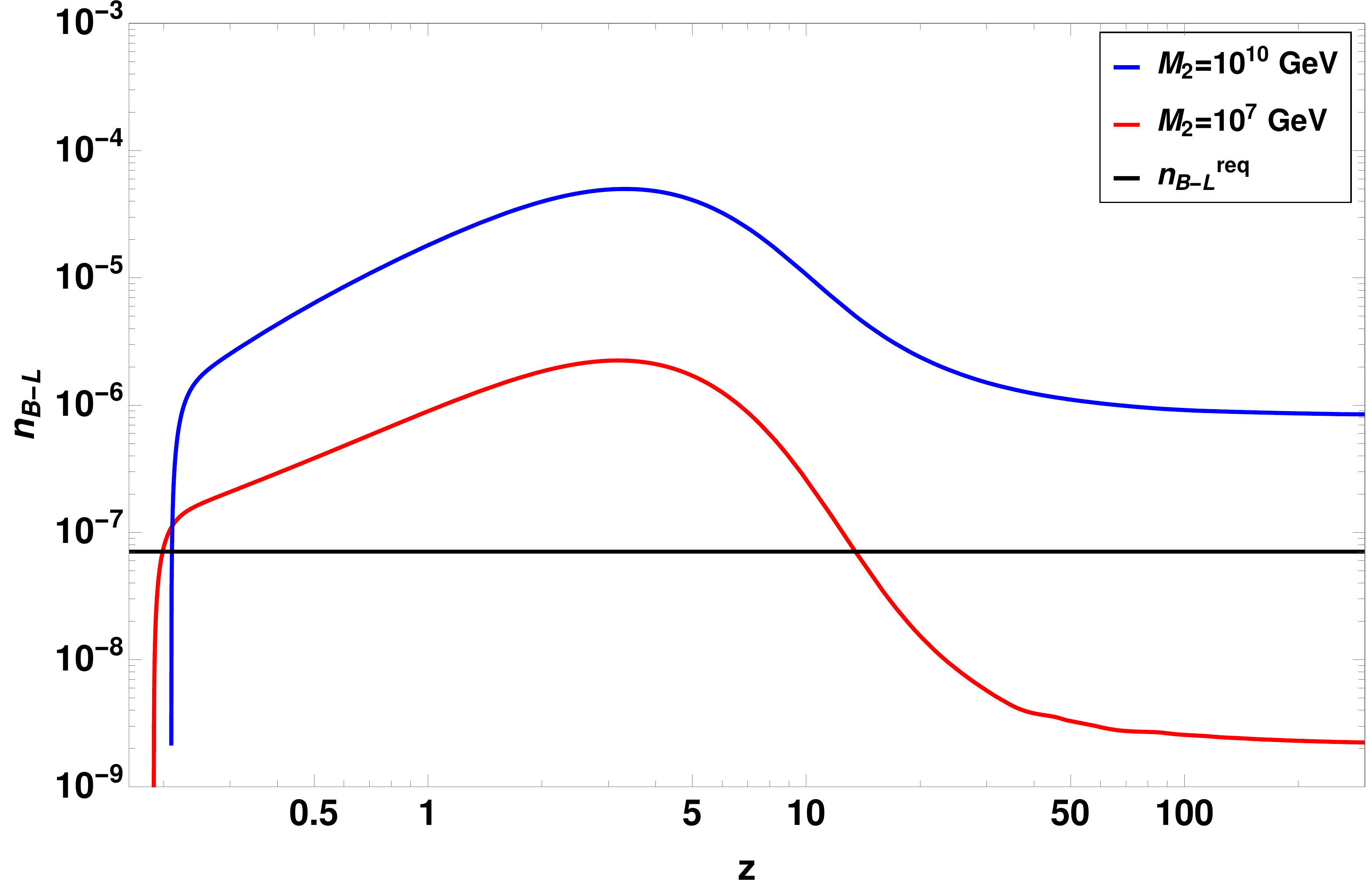}
\caption{Evolution $n_{B-L}$ (Comoving number density of $B-L$) with $z$ for normal ordering. The set of parameters used are: $M_{2}=10^{8}$ GeV, $M_{3}/M_{2}=10^{5}$, $M_{1}=200$ GeV, $m_{\eta}=201$ GeV, $\lambda_{5}=0.05$ for different choices of the lightest neutrino mass $m_{1}$ (Upper left panel); $M_{2}=10^{8}$ GeV, $M_{3}/M_{2}=10^{5}$, $M_{1}=200$ GeV, $m_{\eta}=201$ GeV, $m_1=10^{-5}$ eV for different choices of $\lambda_{5}$ (Upper right panel); $M_{2}=10^{8}$ GeV, $M_{3}/M_{2}=10^{5}$ for $\lambda_{5}=0.05$ with $m_{1}=10^{-5}$ eV for different combinations of $M_{1}$ and $m_{\eta}$ (Bottom left panel); $\lambda_5=0.05, m_{\eta}=201$ GeV,
$M_{1}=200$ GeV, $M_3/M_2=10^{5}, m_{1}=10^{-3}$ eV for different combinations of $M_{2}$ (Bottom right panel). The horizontal black solid line in all the plots indicate the required value of $B-L$ asymmetry to produce the observed baryon asymmetry after sphaleron transitions.}
\label{leptoNH10}
\end{figure}

We first consider normal ordering of light neutrino masses and solve the Boltzmann equations for lepton asymmetry mentioned in the previous section. In order to achieve FIMP type DM so that the Yukawa coupling of $N_1$ comes out to be tiny, we consider the complex matrix $R$ to have the following form  
\[
\textbf{R=}\begin{pmatrix}
1&0&0 \\
0& \cos(z_{R}+iz_{I})& \sin(z_{R}+iz_{I})\\
0& -\sin(z_{R}+iz_{I})& \cos(z_{R}+iz_{I})
\end{pmatrix}
\]
with $z_{R}=0.42$ and $z_{I}=-0.4232$. The justification behind such choice of $R$ and other possibilities of $R$ matrix are mentioned in appendix \ref{appen1}. We further choose the relevant parameters as $m_{\eta}=201$ GeV, $\lambda_{5}=5.0$, $M_{2}=4 \times 10^{7}$ GeV and $\dfrac{M_{3}}{M_{2}} = 10^{5}$ and plot the evolution of comoving number densities of $N_2$ and $N_{B-L}$ as a function of $z$ in figure \ref{leptoNH1}. The final value of $n_{B-L}$ is same as the one required to produce the observed baryon asymmetry after sphaleron transitions. As the temperature cools or $z$ increases, the number density of $N_2$ decreases due to its decay while lepton asymmetry increases. For the chosen benchmark parameters, the washout effects are small or negligible, giving rise to a smooth increase followed by saturation of $n_{B-L}$. To show the significance of washouts, we then choose different benchmark parameters and show the corresponding evolution of $n_{B-L}$ in figure \ref{leptoNH10}. While in the first three plots of this figure, the parameters which are being varied directly affects the washout scatterings, in the fourth plot the change due to variation in $M_2$ is occurring primarily due to change in productions as $M_2$ does not affect the washout scattering processes.

\begin{figure}
\centering
\includegraphics[scale=.22]{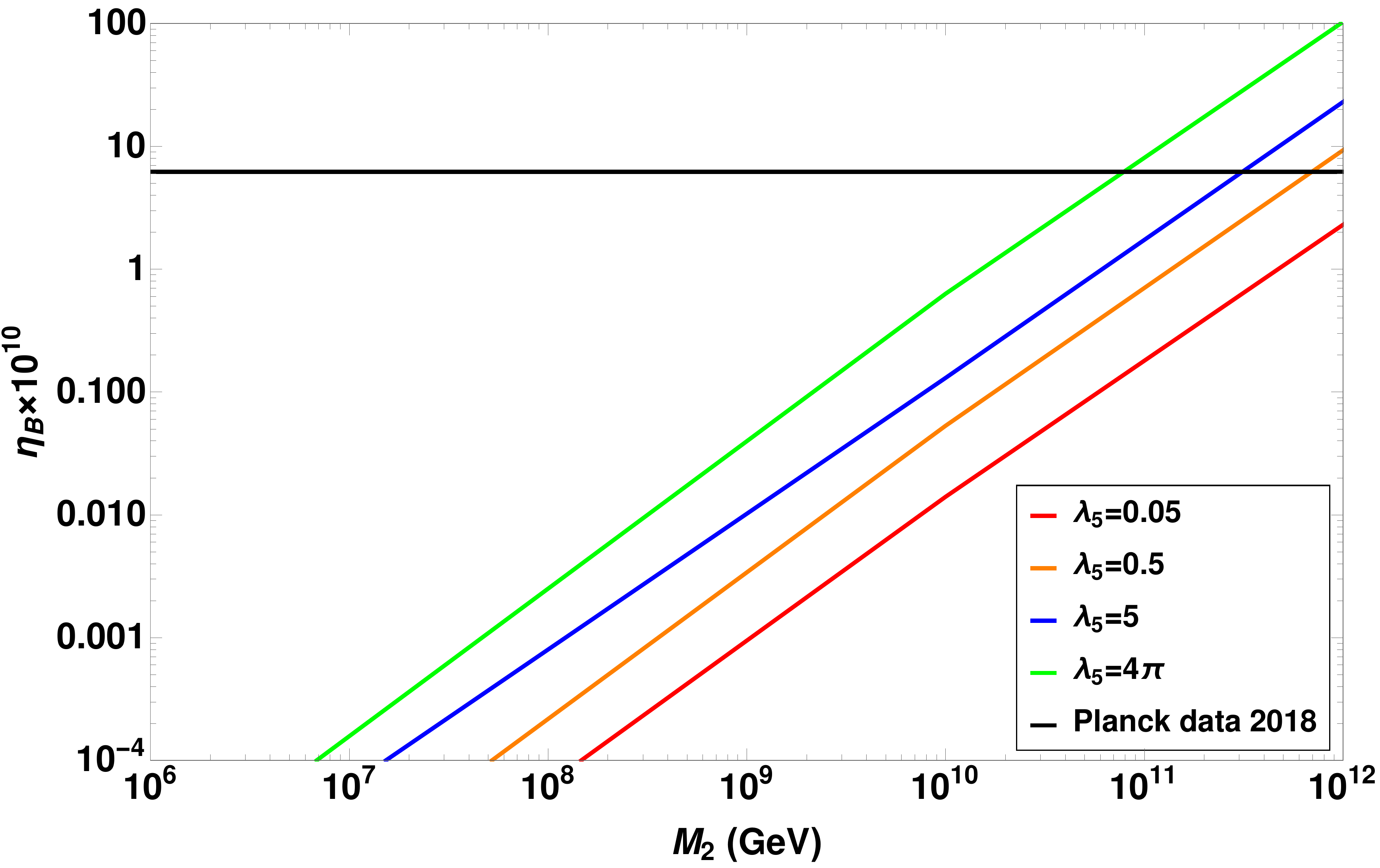}
\includegraphics[scale=.22]{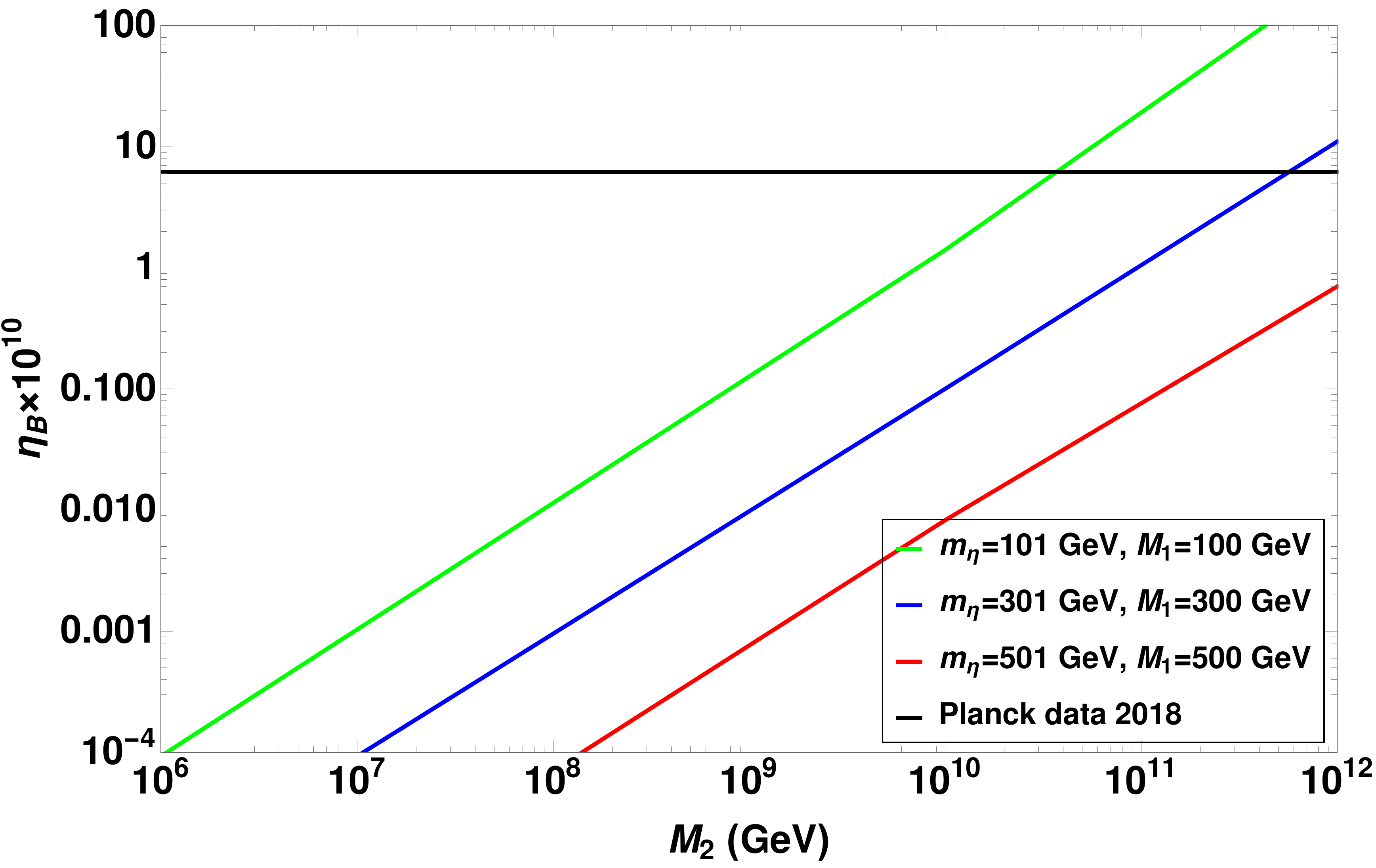} \\
\includegraphics[scale=.22]{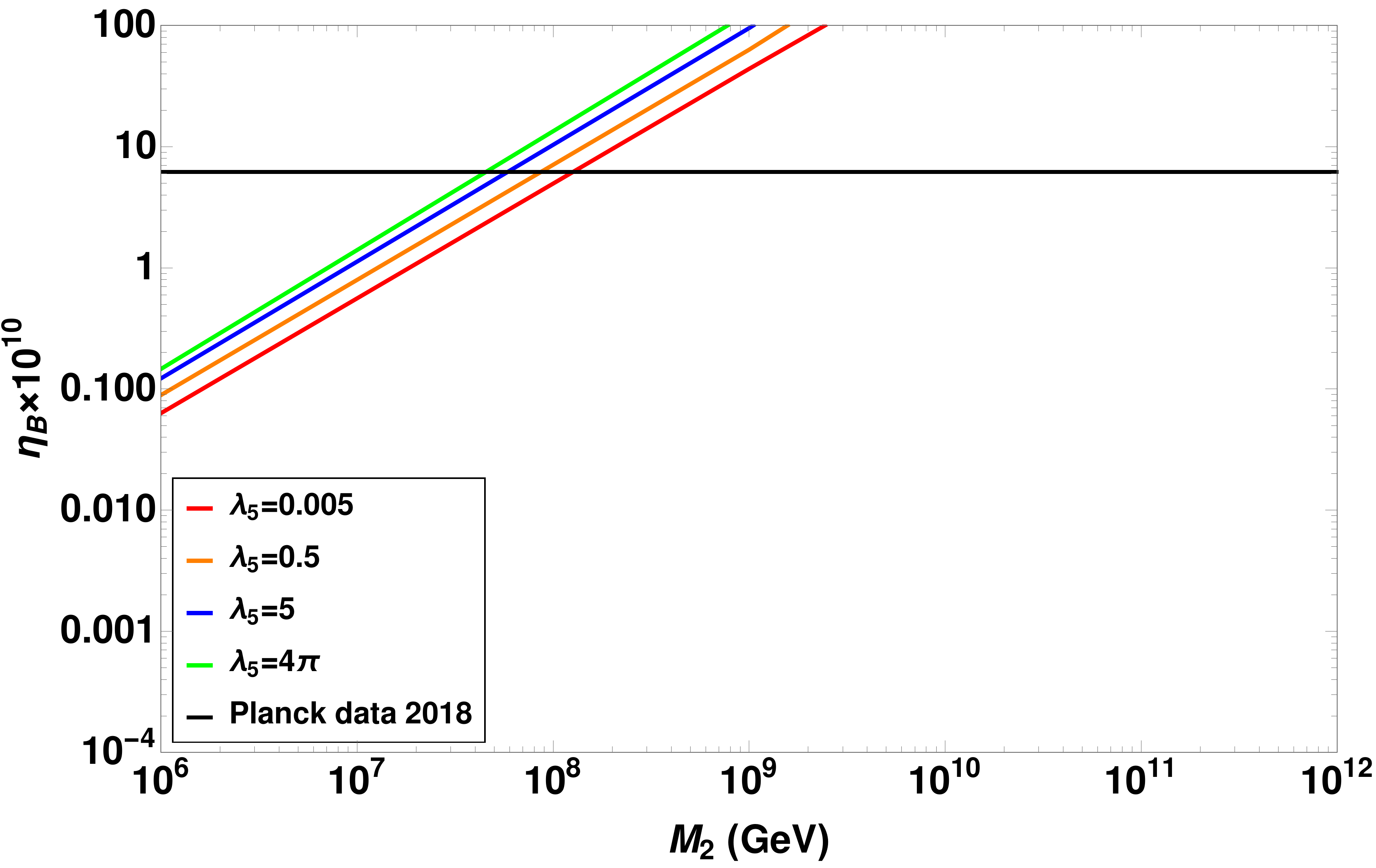}
\includegraphics[scale=.22]{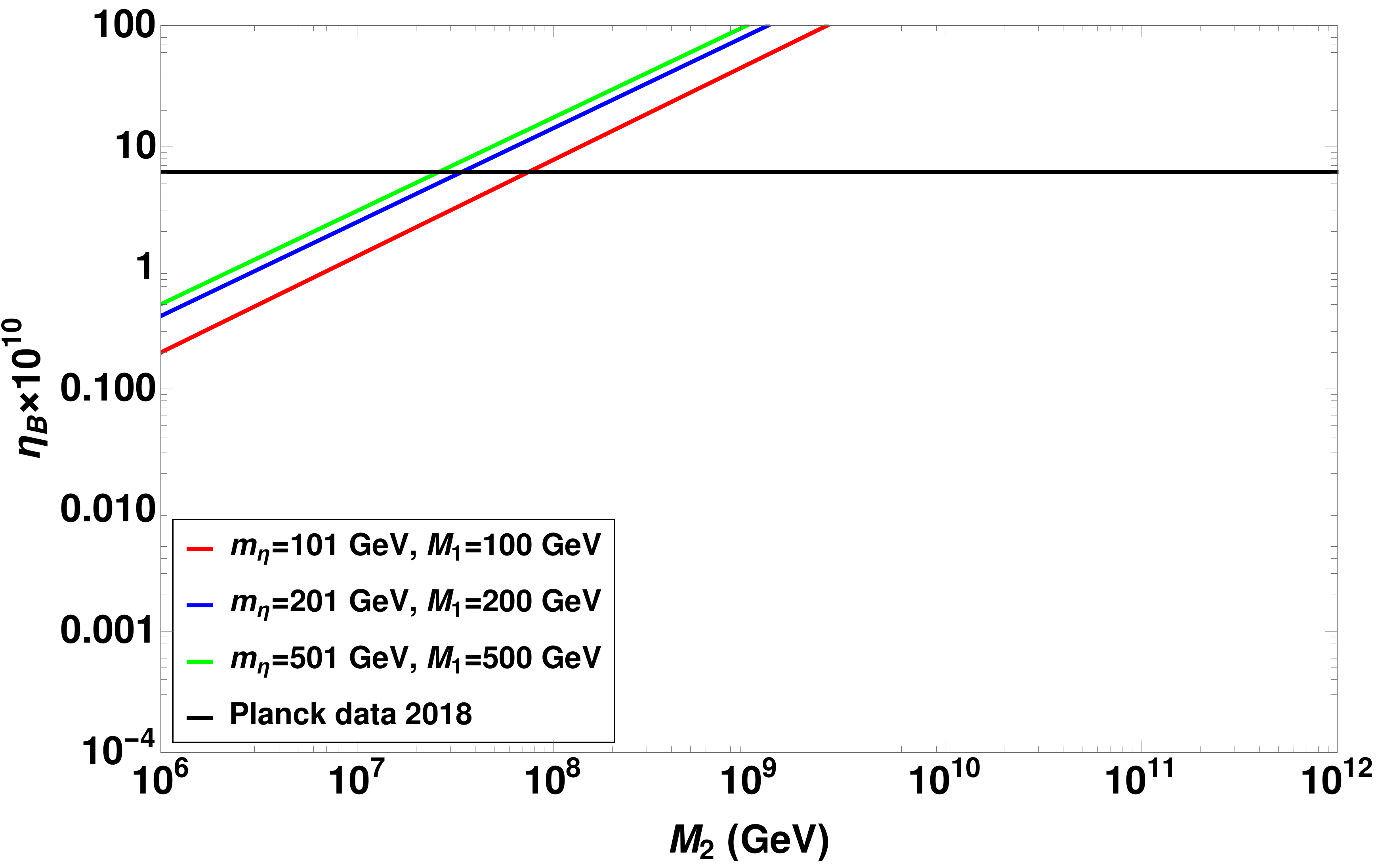}
\caption{Variation of final asymmetry with $M_{2}$ with $m_{1}=10^{-3}$ eV for the set of parameters $m_{\eta}=301$ GeV,  $M_{1}=300$ GeV and $M_{3}/M_{2}=10^{5}$ (Upper left panel); $\lambda_{5}=0.5$ and $M_{3}/M_{2}=10^{5}$ (Upper right panel). Variation of final asymmetry with $M_{2}$ with $m_{1}=10^{-13}$ eV for the set of parameters $m_{\eta}=301$ GeV,  $M_{1}=300$ GeV and $M_{3}/M_{2}=10^{5}$ (Bottom left panel); $\lambda_{5}=0.5$ and $M_{3}/M_{2}=10^{5}$ (Bottom right panel).}
\label{leptoNH2}
\end{figure}

We then evaluate baryon to photon ratio $\eta_B$ from the lepton asymmetry using the formula given in equation \eqref{eq:etaB}. We show the variation of $\eta_B$ with $M_2$ for different benchmark values of relevant parameters like $M_1, m_{\eta}, m_1, \lambda_5$, ratio of heavy singlet fermion masses $M_{3}/M_{2}$ in figure \ref{leptoNH2}. In all these plots, we can see that the correct baryon asymmetry (shown as the horizontal solid black line) can be obtained for different values of these model parameters. As can be seen by comparing the upper panel plots with that in the lower panel, the value of the lightest neutrino mass $m_1$ plays a crucial role in deciding the scale of leptogenesis $M_2$. Lowering the value of $m_1$ leads to lowering of Yukawa couplings which play a role in washout processes (but not in production of lepton asymmetry) thereby lowering the scale of leptogenesis. It is observed that, by playing with the parameters available, the scale of leptogenesis $M_2$ can not be lowered indefinitely, and it remains higher compared to the TeV scale $M_2 \geq 10^7$ GeV.

\begin{figure}
\centering
\includegraphics[scale=.45]{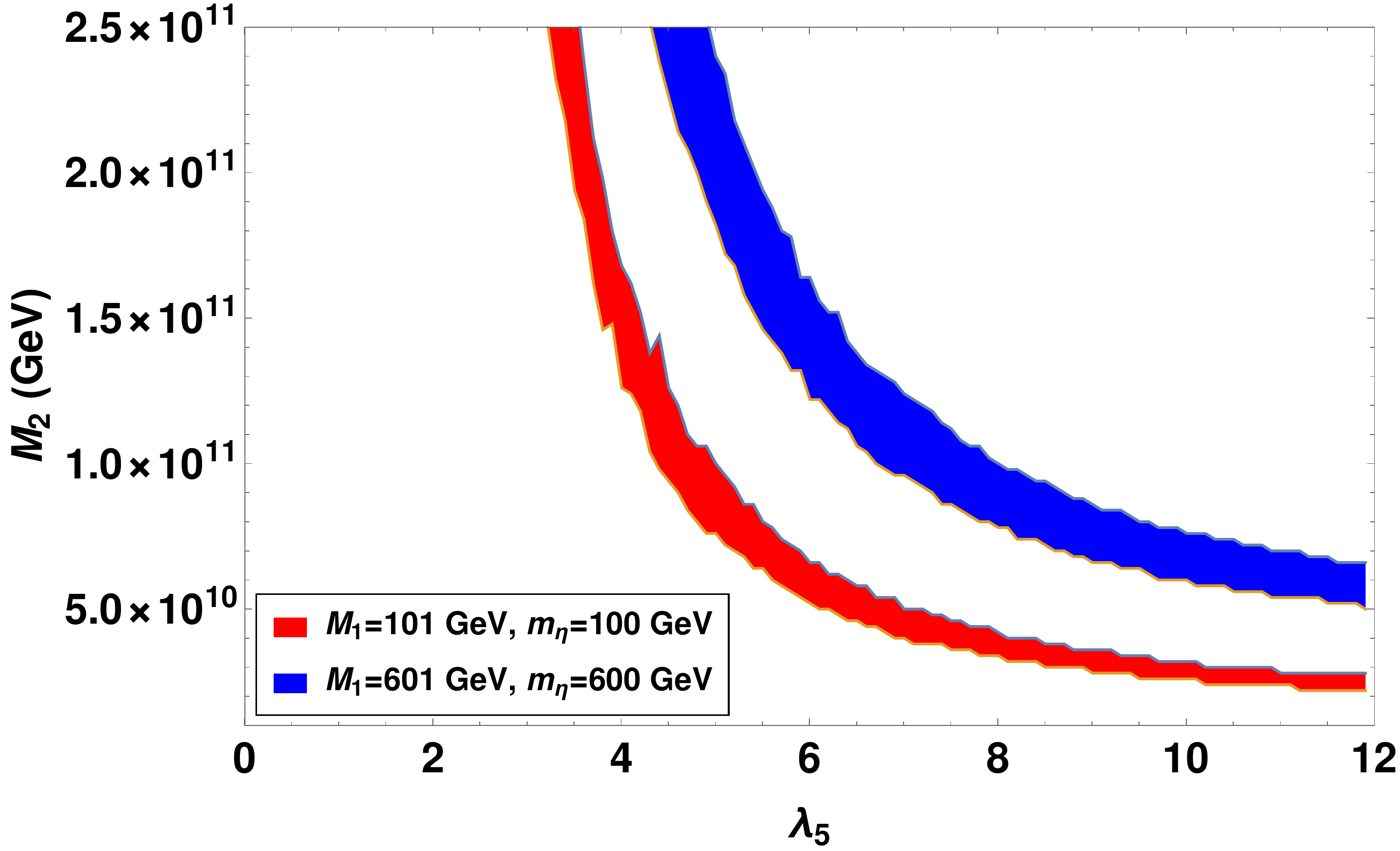}
\caption{Scan plot for $M_{2}$ vs $\lambda_{5}$ for $m_{1}=10^{-3}$ eV with the parameter choice $M_{3}/M_{2}=10^{5}$ for which the observed baryon asymmetry is generated (in case of normal ordering).}
\label{leptoNH5}
\end{figure}

\begin{figure}
\centering
\includegraphics[scale=.45]{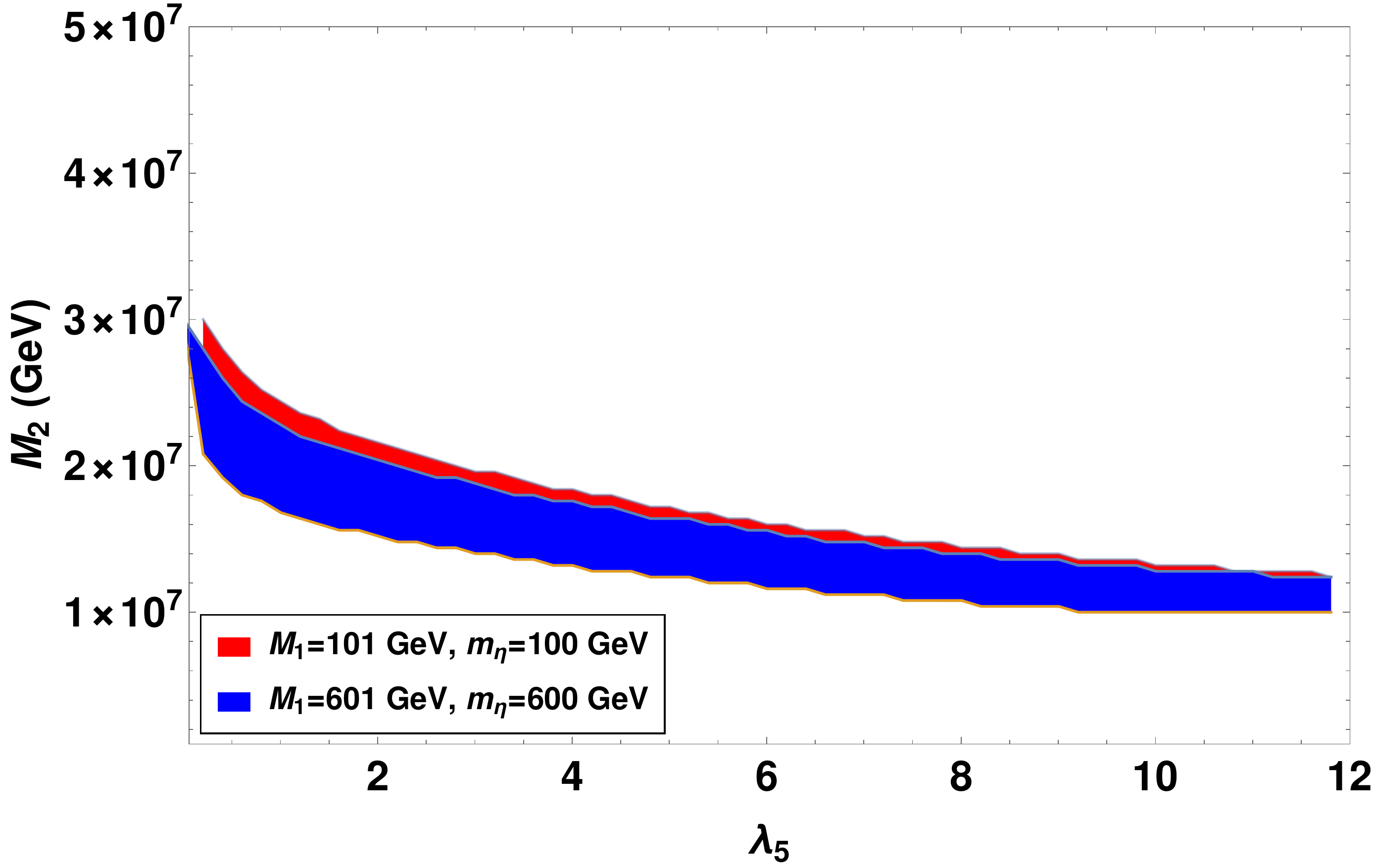}
\caption{Scan plot for $M_{2}$ vs $\lambda_{5}$ for $m_{1}=10^{-13}$ eV with the parameter choice $M_{3}/M_{2}=10^{5}$ for which the observed baryon asymmetry is generated (in case of normal ordering).}
\label{leptoNH5a}
\end{figure}

We finally scan the parameter space in $M_2-\lambda_5$ plane by fixing $M_{3}/M_{2}=10^{5}$ and choosing different combinations of $m_{\eta}, M_1, m_1$. The resulting parameter space that satisfies the correct baryon asymmetry is shown in figure \ref{leptoNH5} and \ref{leptoNH5a} for $m_1=10^{-3}$ eV and $m_1 = 10^{-13}$ eV respectively. Different coloured bands correspond to different combinations of $m_{\eta}, M_1$ as indicated. As can be seen from these plots, the scale of leptogenesis $M_2$ gets pushed up as we decrease $\lambda_5$. Smaller values of $\lambda_5$ result in larger Yukawa couplings from the requirements of light neutrino masses through Casas-Ibarra parametrisation. However, for a fixed lightest neutrino mass and $m_{\eta}, M_1$, such increase in Yukawa couplings will not only enhance the production of asymmetry but also increase the strength of washout scatterings, of which the latter dominates.
For smaller $m_1$, due to the smallness of Yukawa couplings responsible for washout scatterings, the scale of leptogenesis can be slightly lower, as seen by comparing figure \ref{leptoNH5a} with figure \ref{leptoNH5}. In both the cases, we notice that the spectrum of heavy neutrinos $M_{2,3}$ remain very hierarchical, denoting a strong washout regime. To summarise, in case of normal ordering, the scale of leptogenesis can be as low as around $10^7$ GeV, below which successful leptogenesis is not possible.

We then study the possibility of $N_1$ dark matter in the normal ordering scenario of light neutrino masses. To study the FIMP possibility we first choose a benchmark of model parameters which show the possibility of realising tiny Yukawa couplings of $N_1$ required for its non-thermal nature. We fix $M_1 = 300$ GeV, $M_2 = 5.5 \times 10^7$ GeV, $M_3= 10^5 M_2, m_{\eta} = 450$ GeV, $\lambda_5 = 0.1$ and vanishingly small lightest neutrino mass $m_1$. The Dirac Yukawa structure for this choice of benchmark is given, according to Casas-Ibarra parametrisation, by 
\[
\textbf{Y=}\begin{pmatrix}
5.9941\times 10^{-10}+0.i&-0.00030+0.0010i&-6.6655-1.27 40i\\
-7.7134\times 10^{-10}+5.4764\times 10^{-12}i&0.0027-0.0004i&-5.9738+11.7771i\\
6.1755\times 10^{-10}+5.0263\times 10^{-12}i&0.0020-0.0004i&4.3688+8.6048i
\end{pmatrix}
\]
In general, the analytical form of Yukawa matrix, followed from the Casas-Ibarra parametrisation discussed before and for the choice of $R$ matrix mentioned before, can be written as 
\begin{equation}
Y=\begin{pmatrix}
\sqrt{m_{1}}\sqrt{\Lambda_{1}}U_{11}&\sqrt{m_{2}}C^{*}(z)\sqrt{\Lambda_{2}}U_{12}-\sqrt{m_{3}}S^{*}(z)\sqrt{\Lambda_{2}}U_{13}&\sqrt{m_{2}}S^{*}(z)\sqrt{\Lambda_{3}}U_{12}+\sqrt{m_{3}}C^{*}(z)\sqrt{\Lambda_{3}}U_{13} \\
\sqrt{m_{1}}\sqrt{\Lambda_{1}}U_{21}&\sqrt{m_{2}}C^{*}(z)\sqrt{\Lambda_{2}}U_{22}-\sqrt{m_{3}}S^{*}(z)\sqrt{\Lambda_{2}}U_{23}& \sqrt{m_{2}}S^{*}(z)\sqrt{\Lambda_{3}}U_{22}+\sqrt{m_{3}}C^{*}(z)\sqrt{\Lambda_{3}}U_{23}\\
\sqrt{m_{1}}\sqrt{\Lambda_{1}}U_{31}&\sqrt{m_{2}}C^{*}(z)\sqrt{\Lambda_{2}}U_{32}-\sqrt{m_{3}}S^{*}(z)\sqrt{\Lambda_{2}}U_{33}&\sqrt{m_{2}}S^{*}(z)\sqrt{\Lambda_{3}}U_{32}+\sqrt{m_{3}}C^{*}(z)\sqrt{\Lambda_{3}}U_{33}
\end{pmatrix}
\label{yukawamatrix1}
\end{equation}
where $C(z)=\cos{z}$, $S(z)=\sin{z}$ and $U_{ij}$ are the elements of PMNS mixing matrix. Clearly, the Yukawa couplings of $N_1$ to SM leptons and $\eta$ are decided by $m_1$, which is the lightest active neutrino mass in NO. So in case of NO, we can have arbitrarily low Yukawa couplings of $N_1$ required for FIMP dark matter by taking $m_1$ very small. We can also make the Yukawa couplings sizeable by taking $m_1$ similar to the scale of mass squared differences or in the quasi-degenerate light neutrino mass regime, if we want to realise the WIMP scenario for $N_1$. On the contrary, we can not choose $m_1$ to be arbitrarily low for IO of active neutrino mass, thereby restricting our Yukawa couplings to be higher than certain values. However, as we comment in appendix \ref{appen1}, choosing different or more general $R$ matrix can make it possible to have FIMP type Yukawa for $N_1$ in IO which however does not affect the results related to leptogenesis. We will discuss about it later when we go to the discussion of IO part.
\begin{figure}[H]
\centering
\includegraphics[scale=0.23]{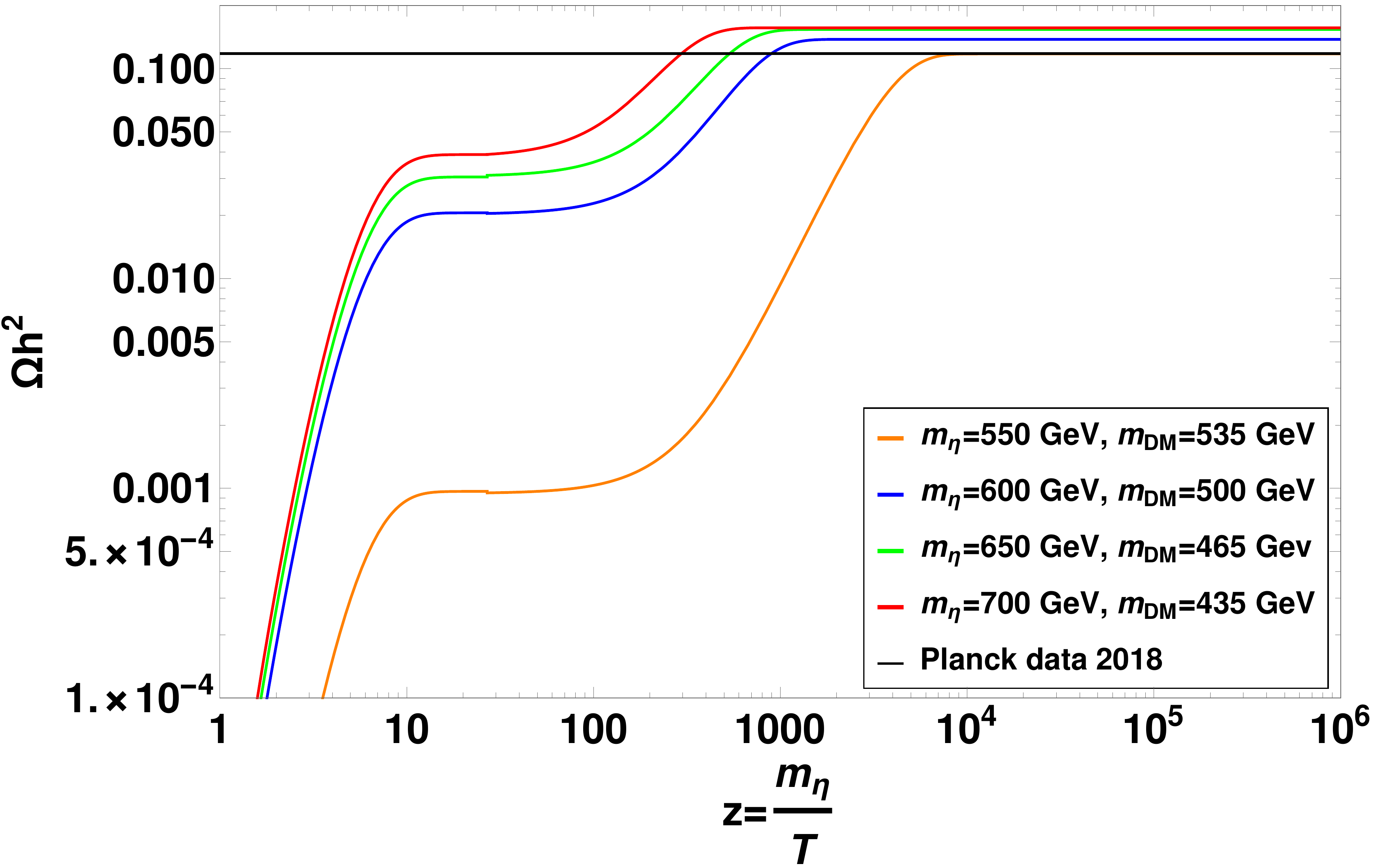}
\includegraphics[scale=.23]{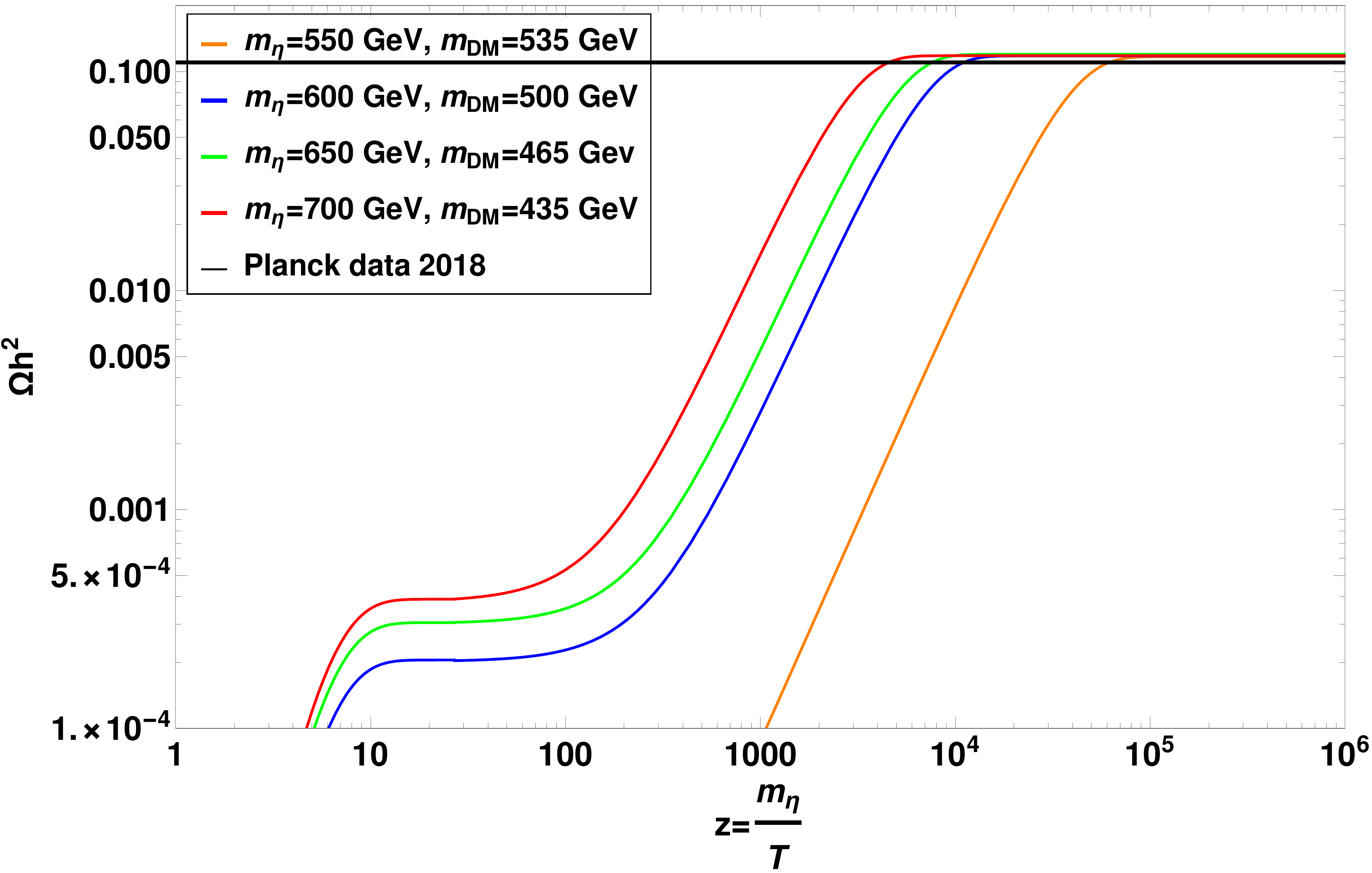}
\caption{Dark Matter ($N_{1}$) relic versus $z=m_{\eta}/T$ taking both equilibrium and out-of-equlibrium contribution. The set of parameters used are $\lambda_{3}+\lambda_{4}+\lambda_{5}=0.001$, $\lambda_{5}=0.1$ and $Y=Y_{11}=Y_{21}=Y_{31}=10^{-9}$ (left panel), $Y=10^{-10}$ (right panel).}
\label{fimpNH1}
\end{figure}

\begin{figure}[H]
\centering
\includegraphics[scale=.6]{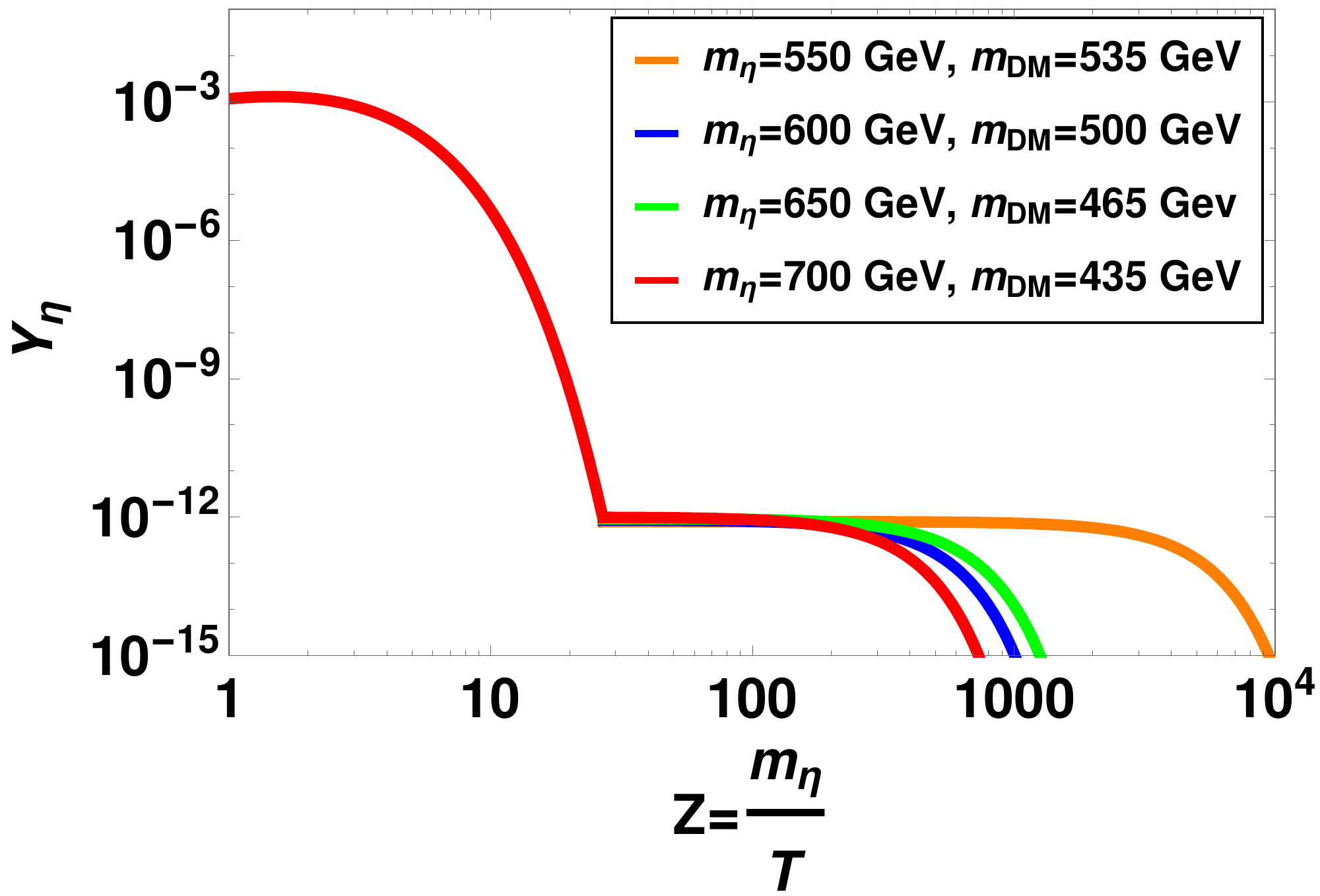}
\caption{Abundance of $\eta$ versus $z=m_{\eta}/T$ for $\lambda_{3}+\lambda_{4}+\lambda_{5}=0.001$, $\lambda_{5}=0.1, Y=10^{-9}.$}
\label{fimpNH2}
\end{figure}

\begin{figure}[H]
\centering
\includegraphics[scale=.4]{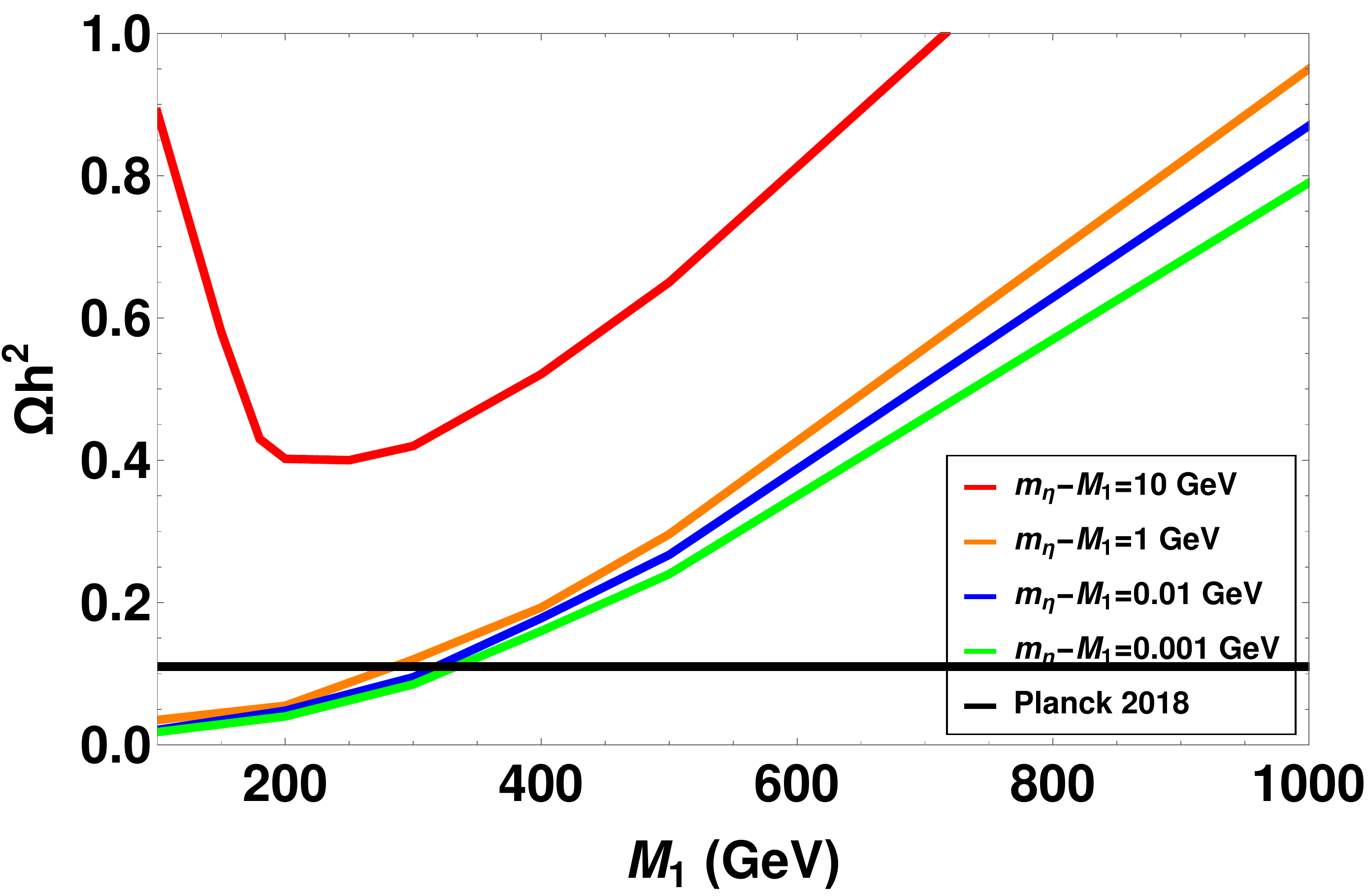}
\caption{WIMP Dark matter relic vs Dark matter mass for different benchmark parameters in case of NO. The chosen benchmark is $\lambda_5=0.005$.}
\label{WIMPNH1}
\end{figure}

To show the FIMP abundance, we choose benchmark values of Yukawa couplings $Y=Y_{11}=Y_{21}=Y_{31}=10^{-10}, 10^{-9}$ for simplicity and show $N_1$ abundance as a function of $z=m_{\eta}/T$ in figure \ref{fimpNH1} for different benchmark values of parameters. As can be seen from these plots, the initial abundance of FIMP is negligible followed by its rise at two distinct epochs: first when the mother particle is in equilibrium and later when the mother particle freezes-out and then decays. Depending upon the Yukawa couplings the equilibrium contribution varies, for example, when the Yukawa coupling is larger the equilibrium contribution to FIMP abundance is also larger. For larger Yukawa, the final abundance of FIMP remains higher, as can be seen by comparing the left and right panel plots of figure \ref{fimpNH1}. The difference due to the choices of $(m_{\eta}, m_{\rm DM})$ is coming as these parameters affect the decay width of $\eta$ into $N_1$. As FIMP mass becomes closer to mother particle's mass, the decay width decreases and hence the yield of DM also decreases slightly. In order to compare the evolution of FIMP abundance with that of mother particle's abundance we also show the variation of $\eta$ abundance as a function of $z$ in figure \ref{fimpNH2}. As can be seen there, the mother particle was in thermal equilibrium in early epochs followed by its thermal freeze-out and then subsequent fall in its abundance at lower temperatures (or higher $z$) due to its decay into FIMP. We have used \texttt{micrOMEGAs} package \cite{Belanger:2013oya} to calculate the freeze-out details of $\eta$ in our work.
\begin{figure}[H]
\centering
\includegraphics[scale=.6]{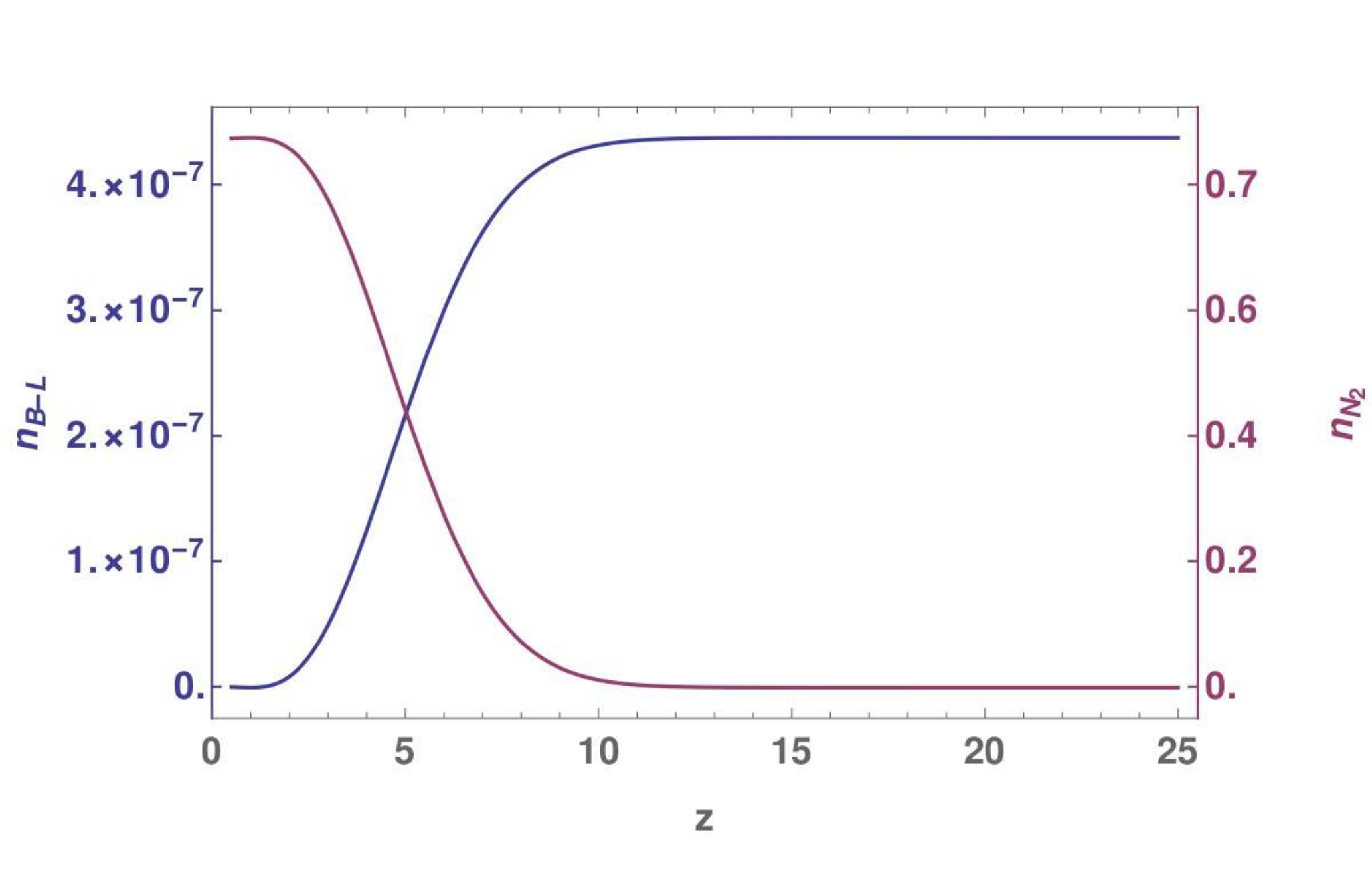}
\caption{Evolution $n_{N_2}, n_{B-L}$ (Comoving number densities of of $N_2$, $B-L$) with $z$ for inverted ordering. The set of parameters used are $M_{2}=4 \times 10^{4}$ GeV, $M_{3}/M_{2}=10$, $M_{1}=201$ GeV, $m_{\eta}=201$ GeV, $\lambda_{5}=10^{-4}, m_3 = 10^{-13}$ eV.}
\label{leptoIH01}
\end{figure}

We also check the possibility of $N_1$ as WIMP DM in NO case. However, for WIMP DM we need much larger Yukawa couplings than the ones mentioned above for FIMP. Such larger couplings are required in order to produce $N_1$ thermally in the early universe which later undergoes thermal freeze-out leaving the right relic abundance. We generate such large Yukawa couplings by increasing the lightest active neutrino mass to $m_{1}=10^{-2}$ eV from $m_{1}=10^{-13}$ eV before. Such increase in lightest active neutrino mass however, crucially affects the leptogenesis results for NO, shifting the scale of leptogenesis $M_2$, which we discussed earlier. The relic abundance for WIMP DM as a function of its mass is shown in figure \ref{WIMPNH1} for different benchmark parameters. As can be seen from this plot, the mass splitting between $\eta$ and $N_1$ plays a crucial role in generating the correct abundance. For smaller mass splittings the coannihilation between $\eta$ and $N_1$ gets enhanced, bringing down the relic abundance within the observed limits. Here also we have implemented the model in \texttt{micrOMEGAs} to calculate the relic abundance of $N_1$.

\begin{figure}
\centering
\includegraphics[scale=.23]{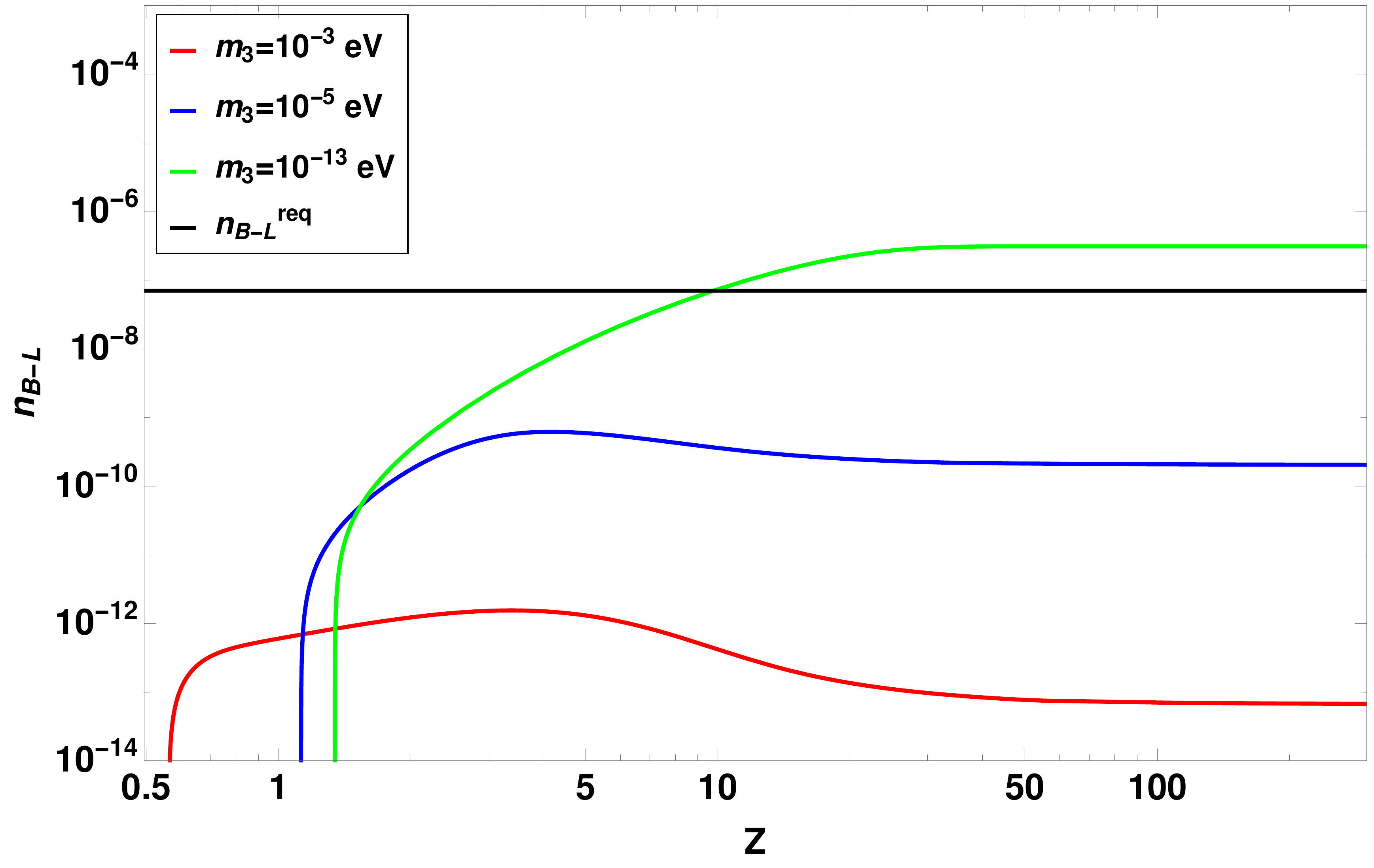}
\includegraphics[scale=.23]{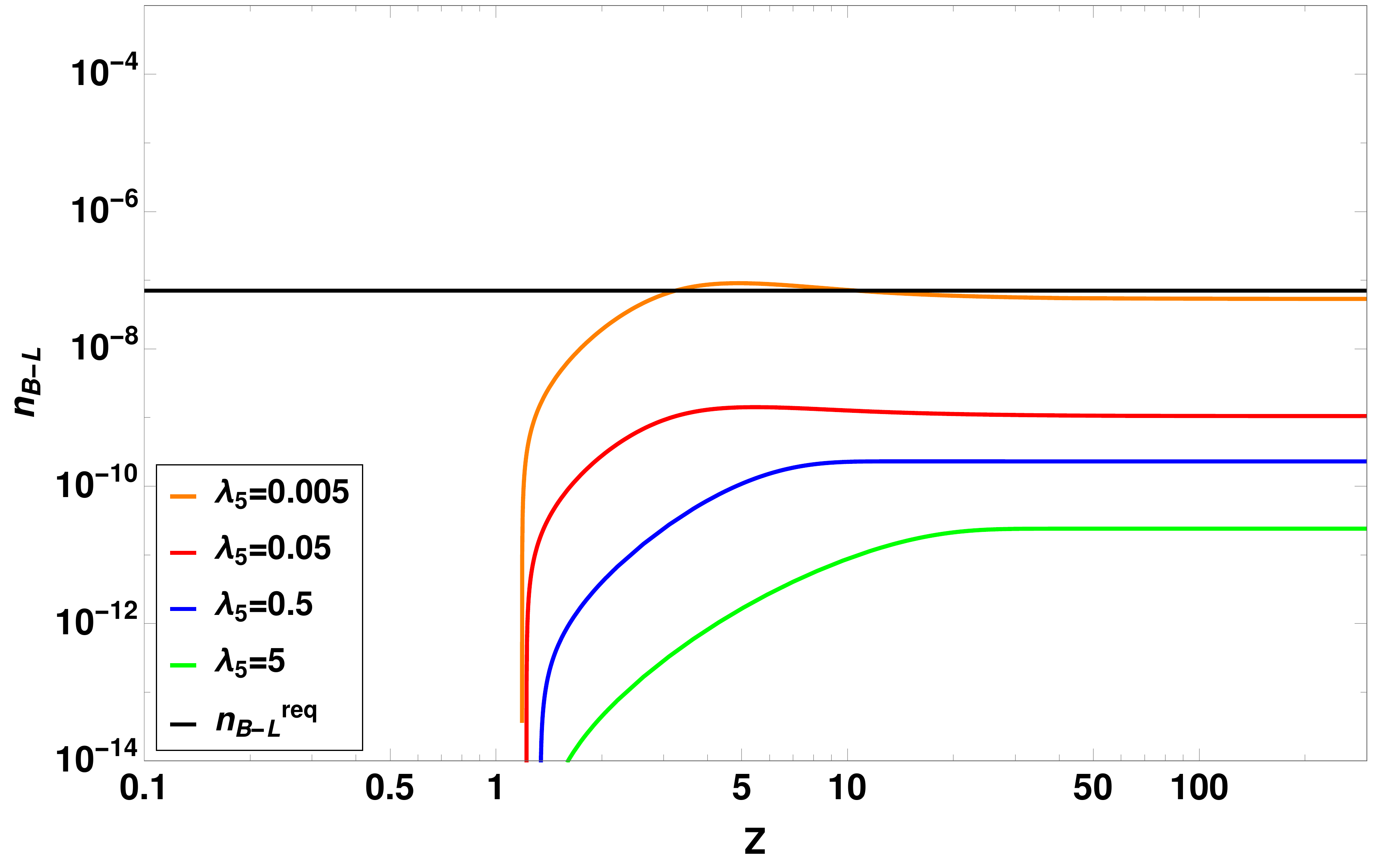} \\
\includegraphics[scale=.23]{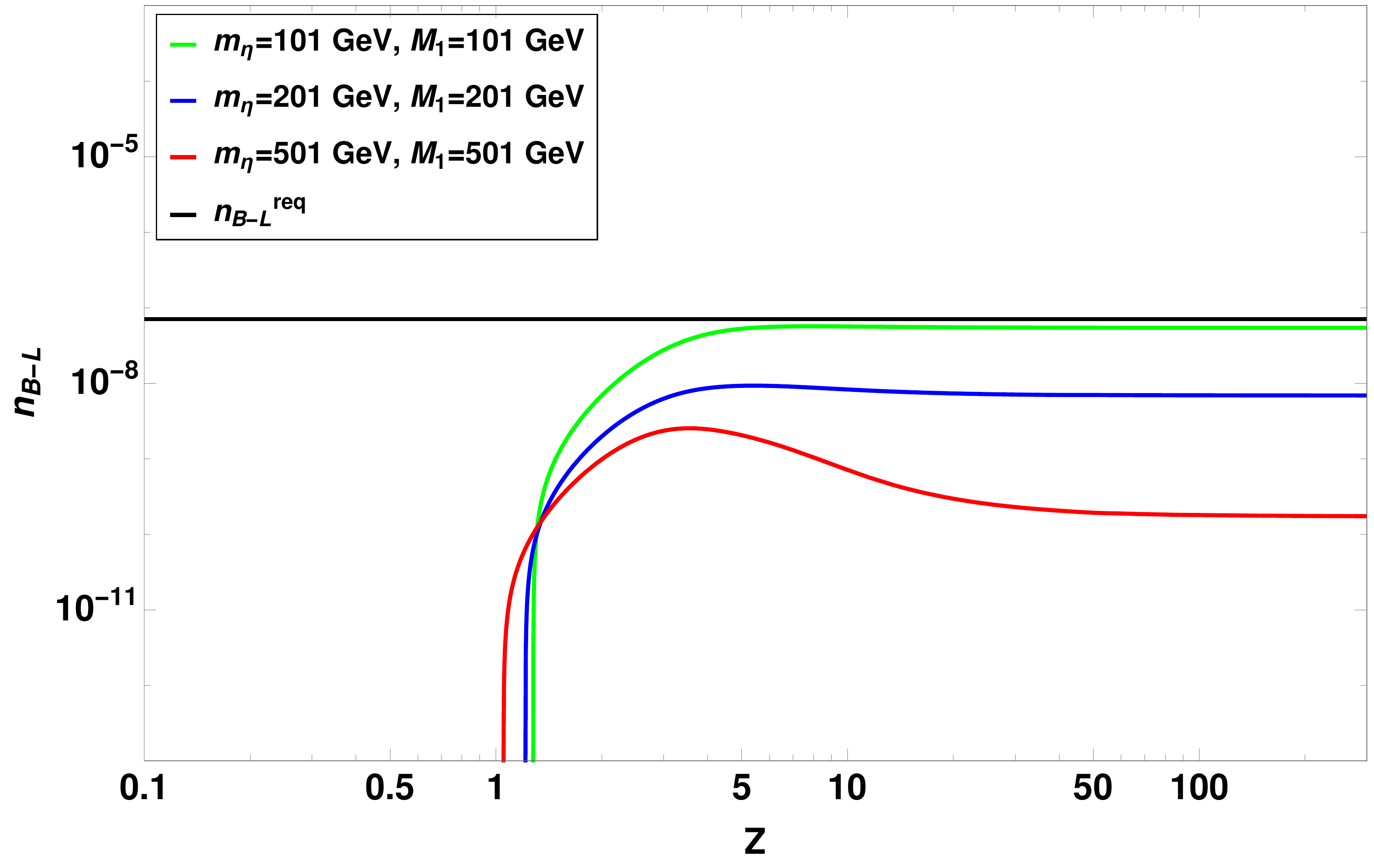}
\includegraphics[scale=.23]{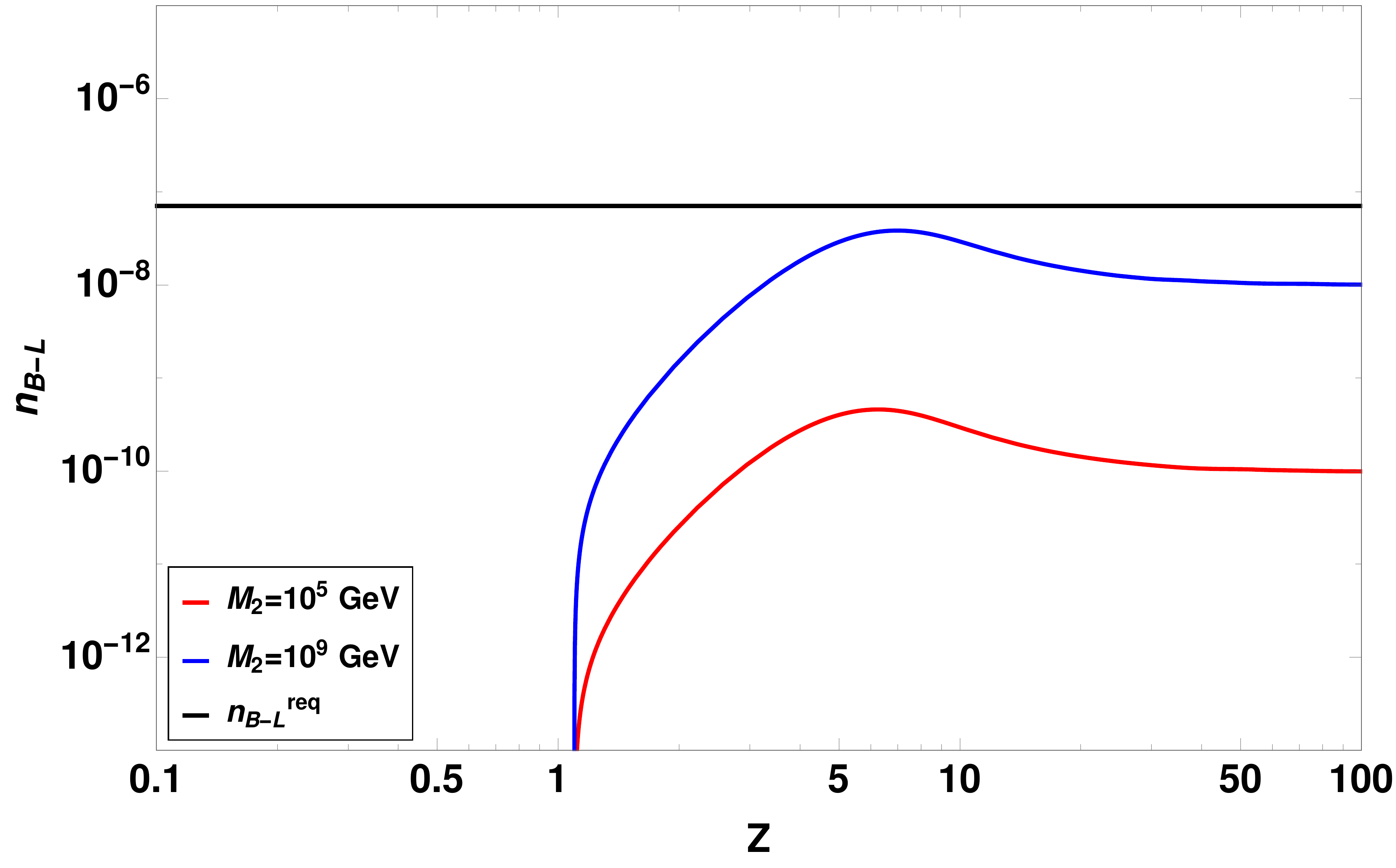}
\caption{Evolution $n_{B-L}$ (Comoving number density of $B-L$) with $z$ for inverted ordering. The set of parameters used are: $M_{2}=10^{5}$ GeV, $M_{3}/M_{2}=10^{5}$, $M_{1}=200$ GeV, $m_{\eta}=201$ GeV, $\lambda_{5}=0.5$ for different choices of the lightest neutrino mass $m_{3}$ (Upper left panel); $M_{2}=10^{5}$ GeV, $M_{3}/M_{2}=10^{5}$, $M_{1}=200$ GeV, $m_{\eta}=201$ GeV, $m_3=10^{-5}$ eV for different choices of $\lambda_{5}$ (Upper right panel); $M_{2}=10^{5}$ GeV, $M_{3}/M_{2}=10^{5}$ for $\lambda_{5}=0.5$ with $m_{3}=10^{-5}$ eV for different combinations of $M_{1}$ and $m_{\eta}$ (Bottom left panel); $m_{\eta}=201, M_{1}=200, \lambda_5=0.05, M_3/M_2=10^{3},
m_{3}=10^{-3}$ eV for different combinations of $M_{2}$ (Bottom right panel). The horizontal black solid line in all the plots indicate the required value of $B-L$ asymmetry to produce the observed baryon asymmetry after sphaleron transitions.}
\label{leptoIH0}
\end{figure}

We now move onto discussing the results for inverted ordering of light neutrino masses. We choose the following R matrix in order to generate the desired Yukawa structure:
\[
\textbf{R=}\begin{pmatrix}
1&0&0 \\
0& \cos{(z_{R}+iz_{I})} & \sin{(z_{R}+iz_{I})} \\
0 & -\sin{(z_{R}+iz_{I})} & \cos{(z_{R}+iz_{I})}
\end{pmatrix}
\]
with $z_{R}=1.5707$ and $z_{I}=-0.0008$. Once again, the justification behind such choice of $R$ and other possibilities of $R$ matrix are mentioned in appendix \ref{appen1}. The following Yukawa structure is obtained for $N_2$ leptogenesis with the benchmark parameter as $M_{1}=300$ GeV, $M_{2}=5.5\times 10^{4}$ GeV, 
$M_{3}/M_{2}=10$, $m_{\eta}=450$ GeV, $\lambda_{5}=0.001$ and vanishingly small lightest neutrino mass $m_3$.
\[
\textbf{Y=}\begin{pmatrix}
0.00042+0.i&2.011\times 10^{-8}-3.243\times 10^{-8}i&0.00171+9.587\times 10^{-13}i\\
-0.00026+0.00003i&-1.3643\times 10^{-7}-9.104\times 10^{-9}&0.000222-0.000159i\\
0.00027+0.00003i&-1.413\times 10^{-7}+8.322\times 10^{-9}i&-0.00215+0.000149i
\end{pmatrix}
\]
Clearly, the Yukawa coupling of the lightest right handed neutrino do not become arbitrarily small, unlike in NO case mentioned earlier.

\begin{figure}
\centering
\includegraphics[scale=.3]{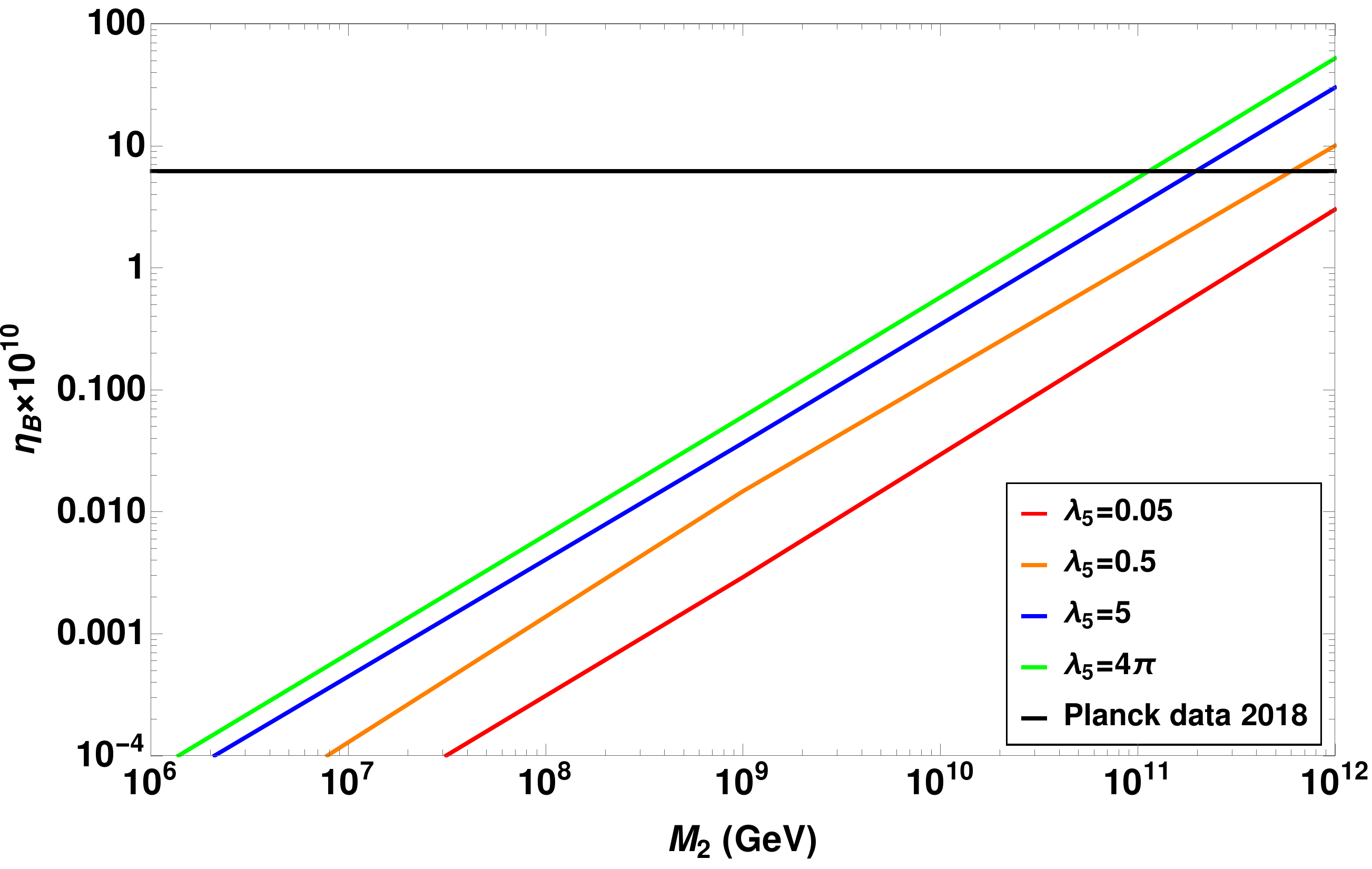}
\includegraphics[scale=.3]{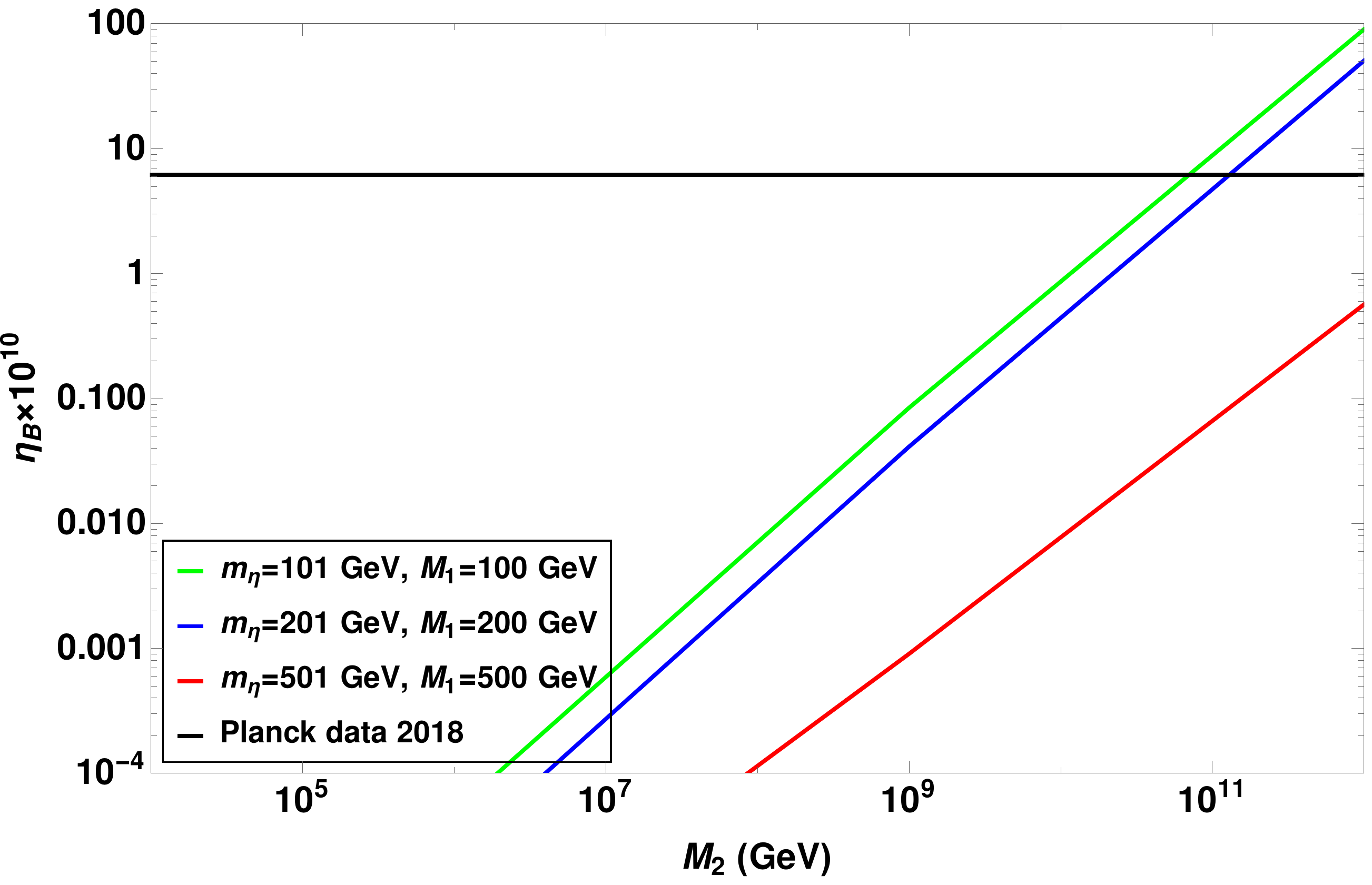} \\
\includegraphics[scale=.3]{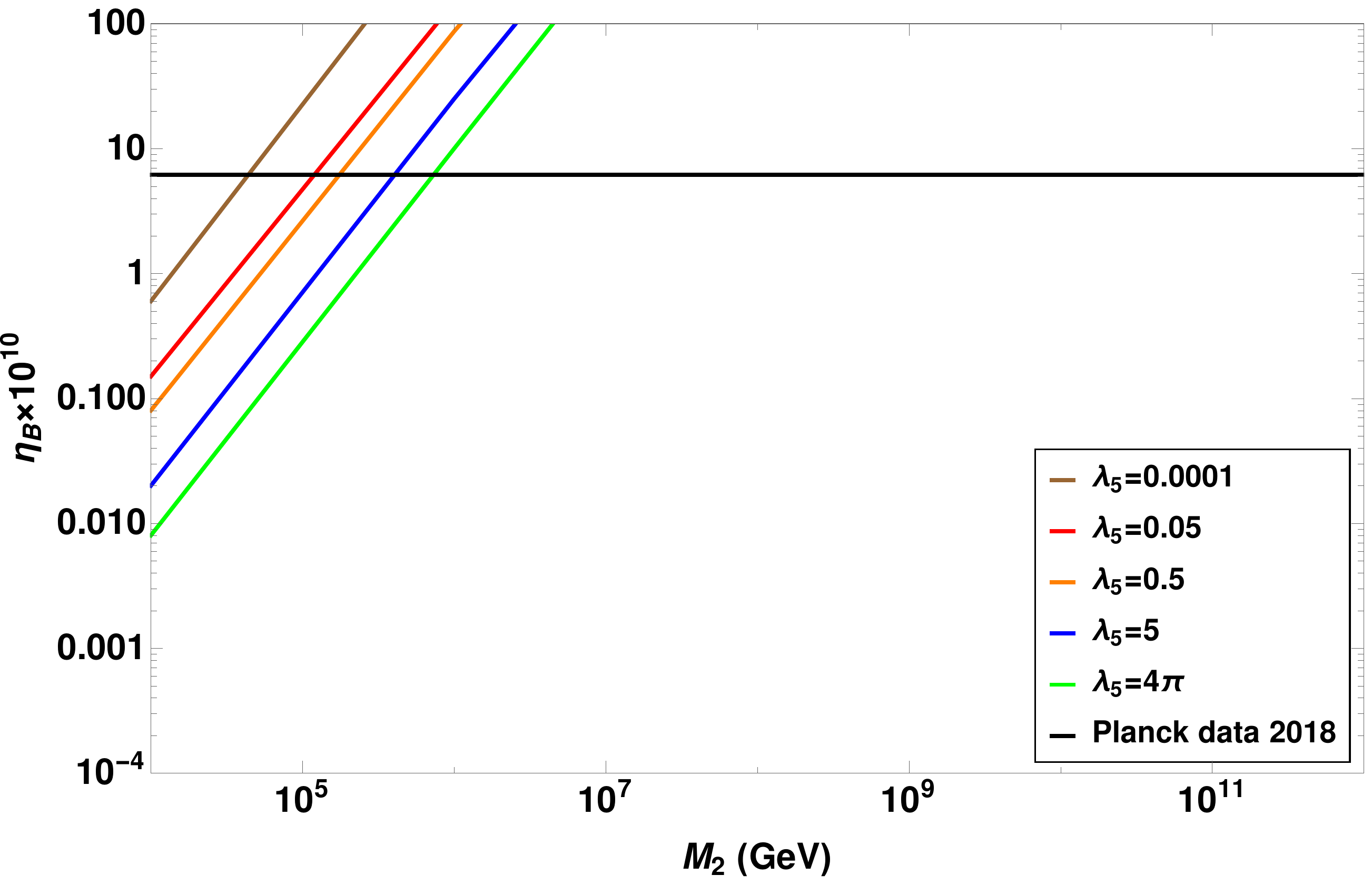}
\includegraphics[scale=.23]{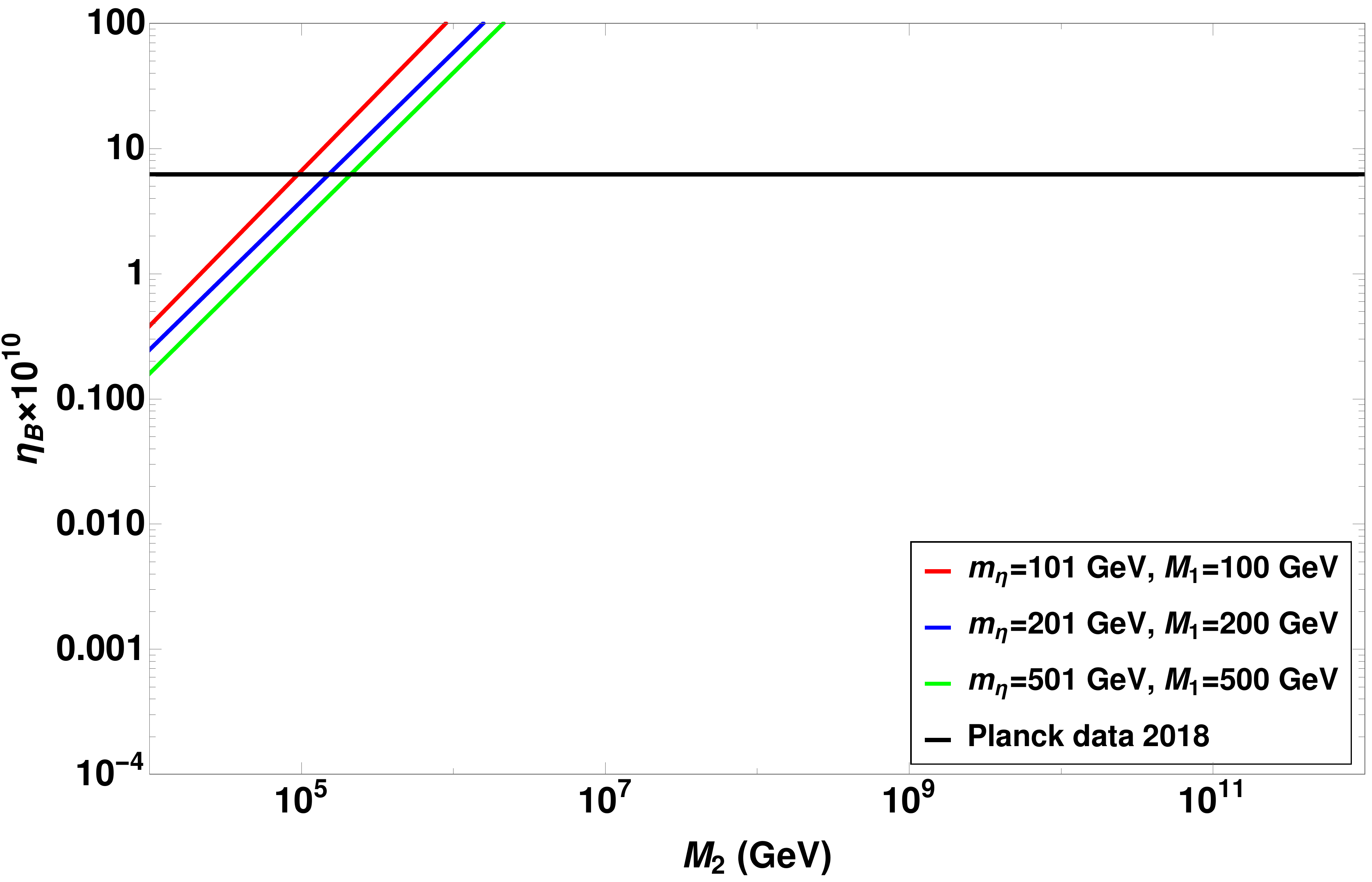}
\caption{Variation of final asymmetry with $M_{2}$ with $m_{3}=10^{-3}$ eV for the set of parameters $m_{\eta}=301$ GeV,  $M_{1}=300$ GeV and $M_{3}/M_{2}=10^{5}$ (Upper left panel); $\lambda_{5}=0.5$,  $M_{3}/M_{2}=10^{5}$ (Upper right panel). Variation of the final asymmetry with $M_{2}$ for $m_{3}=10^{-13}$ eV for the set of parameters $m_{\eta}=301$ GeV,  $M_{1}=300$ GeV and $M_{3}/M_{2}=10^{2}$ (Bottom left panel); $\lambda_{5}=0.5$, $M_{3}/M_{2}=10^{2}$ (Bottom right panel).}
\label{leptoIH1}
\end{figure}

In figure \ref{leptoIH01}, we show the evolution of comoving number densities for $N_2$ and $B-L$ asymmetry for chosen benchmark parameters $M_{2}=4 \times 10^{4}$ GeV, $M_{3}/M_{2}=10$, $M_{1}=201$ GeV, $m_{\eta}=201$ GeV, $\lambda_{5}=10^{-4}, m_3 = 10^{-13}$ eV. Clearly, the number density of $N_2$ decreases due to its decay while the $B-L$ asymmetry increases as $N_2$ abundance decreases. Once again, the final value of $n_{B-L}$ satisfies the criteria of producing correct baryon asymmetry after sphaleron transitions and the washout effects are minimal for the chosen parameters. To show the importance of washout effects, we then choose different benchmark parameters relevant to the production as well as washout of lepton asymmetry and show the corresponding evolution of $n_{B-L}$ in figure \ref{leptoIH0}. Compared to the NO case shown in figure \ref{leptoNH10}, here the washout effects remain slightly smaller, but still can be very relevant.

\begin{figure}
\centering
\includegraphics[scale=.4]{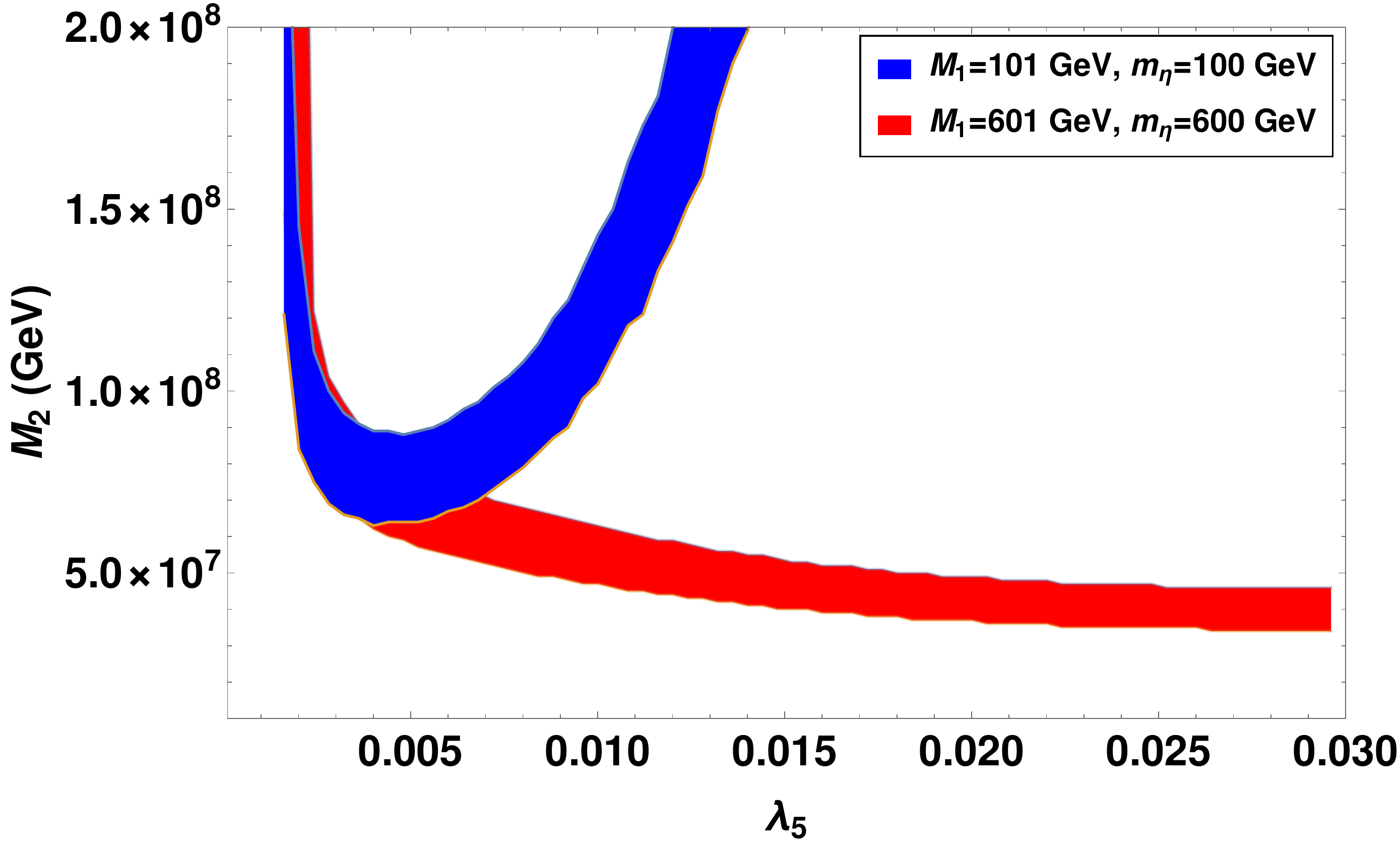}
\caption{Parameter space in $M_{2}$ vs $\lambda_{5}$ plane for $m_{3}=10^{-3}$ eV with the parameter choice $M_{3}/M_{2}=10^{5}$ for which the observed baryon asymmetry is generated for inverted ordering.}
\label{leptoscanIH1}
\end{figure}

\begin{figure}
\centering
\includegraphics[scale=.4]{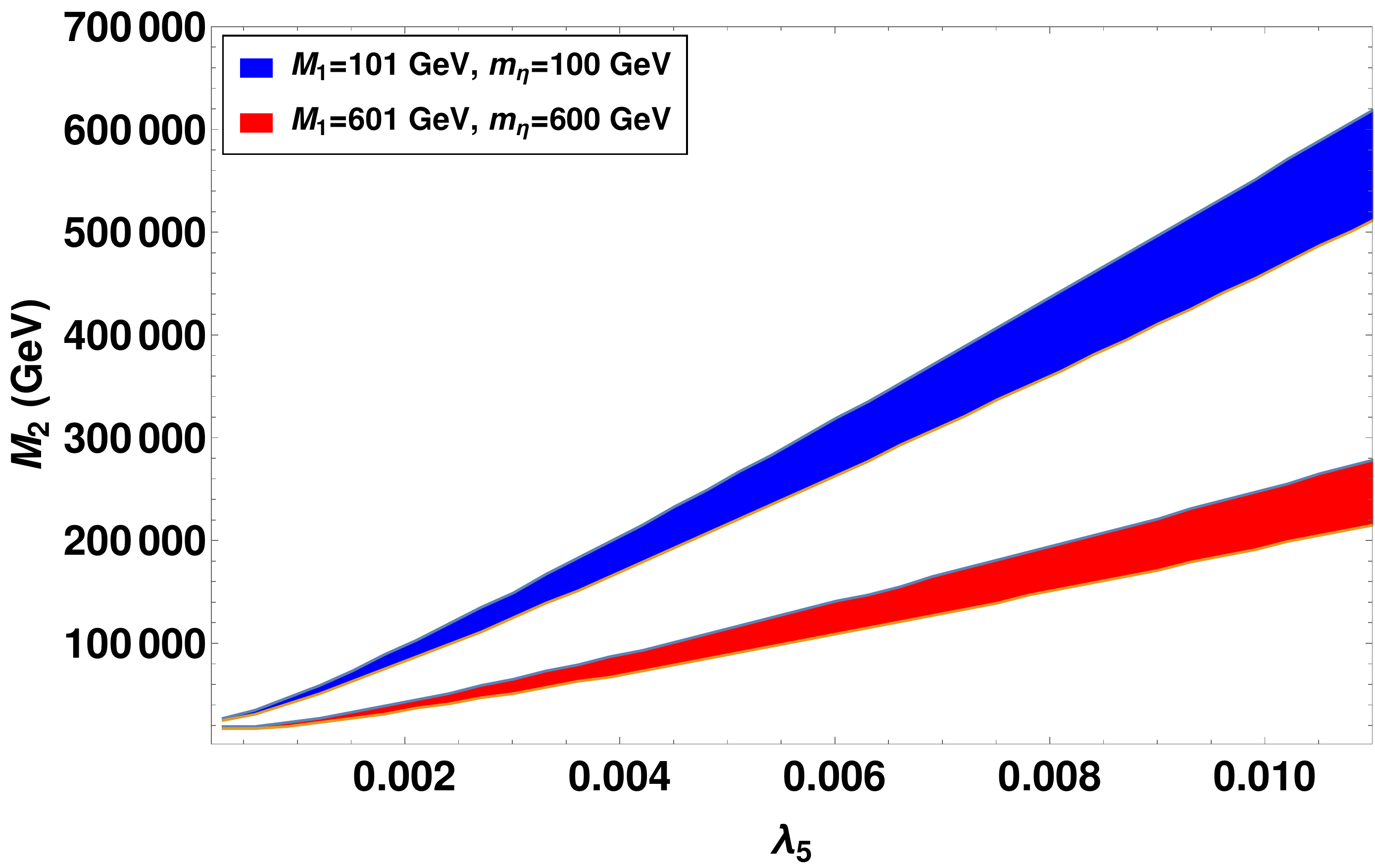}
\caption{Parameter space in $M_{2}$ vs $\lambda_{5}$ plane for $m_{3}=10^{-13}$ eV with the parameter choice $M_{3}/M_{2}=10^{2}$ for which the observed baryon asymmetry is generated for inverted ordering.}
\label{leptoscanIH1a}
\end{figure}

We then show the variation of baryon to photon ratio with mass of $N_2$ for different benchmark parameters in figure \ref{leptoIH1} and compare it with the observed baryon asymmetry. Clearly, the observed baryon asymmetry can be produced by appropriate choices of benchmark parameters. Interestingly, the scale of leptogenesis can be as low as few tens of TeV, unlike in case of NO where the scale of leptogenesis was several order of magnitudes above TeV scale. This is particularly possible when both $\lambda_5$ and lightest neutrino mass $m_3$ are chosen small and we are in a weak washout regime. In such a case the heavy right handed neutrinos become less hierarchical as well. We finally show the parameter space in terms of $M_{2}$ and $\lambda_5$ which leads to the observed baryon asymmetry for different choices of $m_{\eta}, M_1, m_3$ and show them in figure \ref{leptoscanIH1} and figure \ref{leptoscanIH1a} for $m_3=10^{3}$ eV and $m_3=10^{-13}$ eV respectively. Sharp contrast can be noticed from similar scan plots for normal ordering shown in figure \ref{leptoNH5} and \ref{leptoNH5a}. Only for a particular benchmark choice of $m_{\eta}=600$ GeV, $M_1=601$ GeV, $m_3=10^{-3}$ eV, we see that the scale of leptogenesis is pushed to higher values as we decrease $\lambda_5$. For a different combination of $m_{\eta}, M_1$ but with same $m_3$, the scale of leptogenesis decreases first, as we decrease $\lambda_5$, but at some point, it again rises. This is due to the similar reason we mentioned in the discussion of NO results that washout effects start dominating over the production processes as we lower $\lambda_5$ further leading to corresponding rise in Yukawa couplings. For very small value of the lightest neutrino mass $m_3=10^{-13}$ eV, the washout effects are negligible and the scale of leptogenesis can be lowered down to few tens of TeV, as seen from the plot in figure \ref{leptoscanIH1a}. Due to such weak washout regime for $m_3=10^{-13}$ eV, the hierarchy between heavy neutrinos $M_{2,3}$ can be lowered down as well, as can be seen from the plot in figure \ref{leptoscanIH1a}.

\begin{figure}
\centering
\includegraphics[scale=.22]{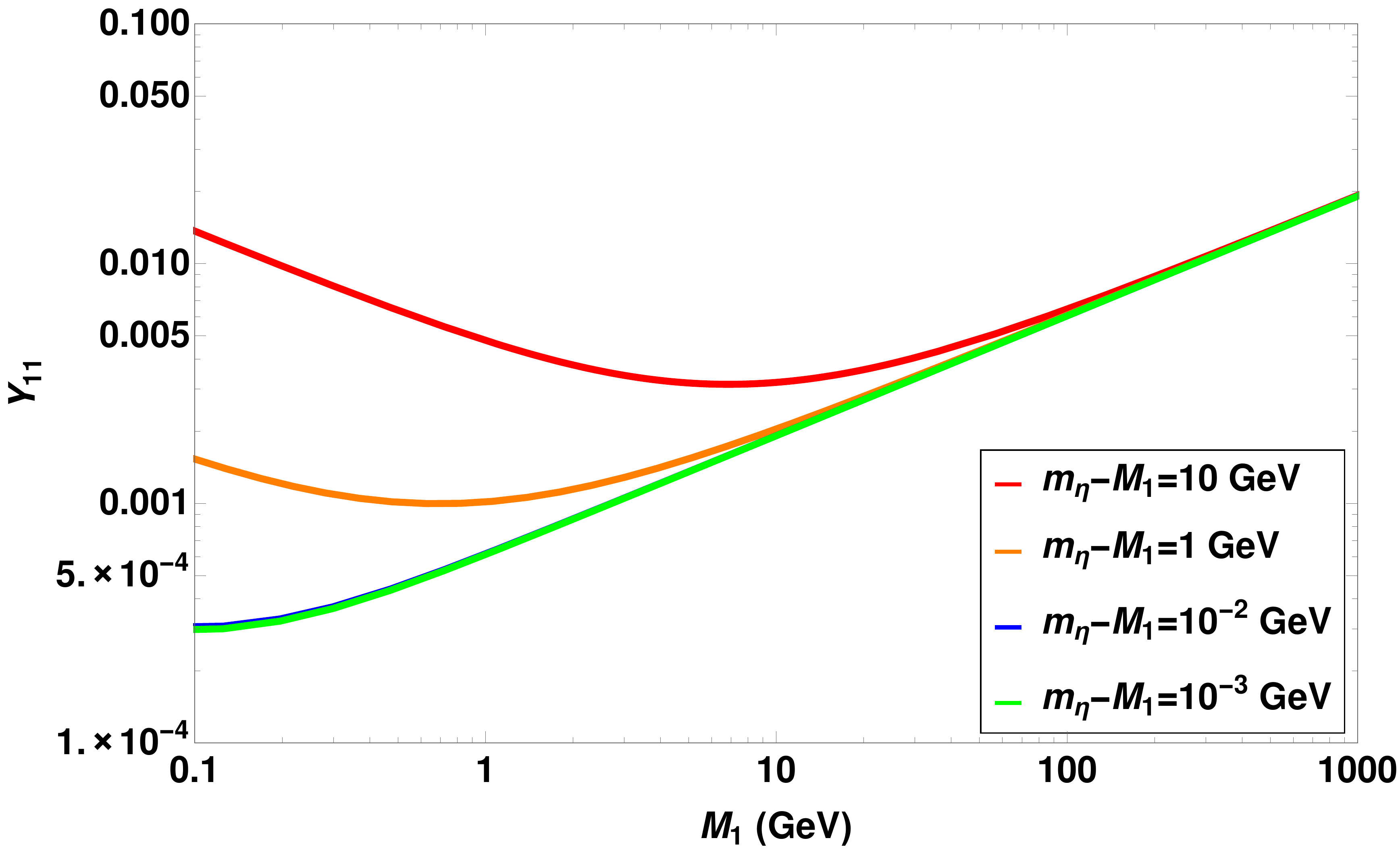}
\includegraphics[scale=.22]{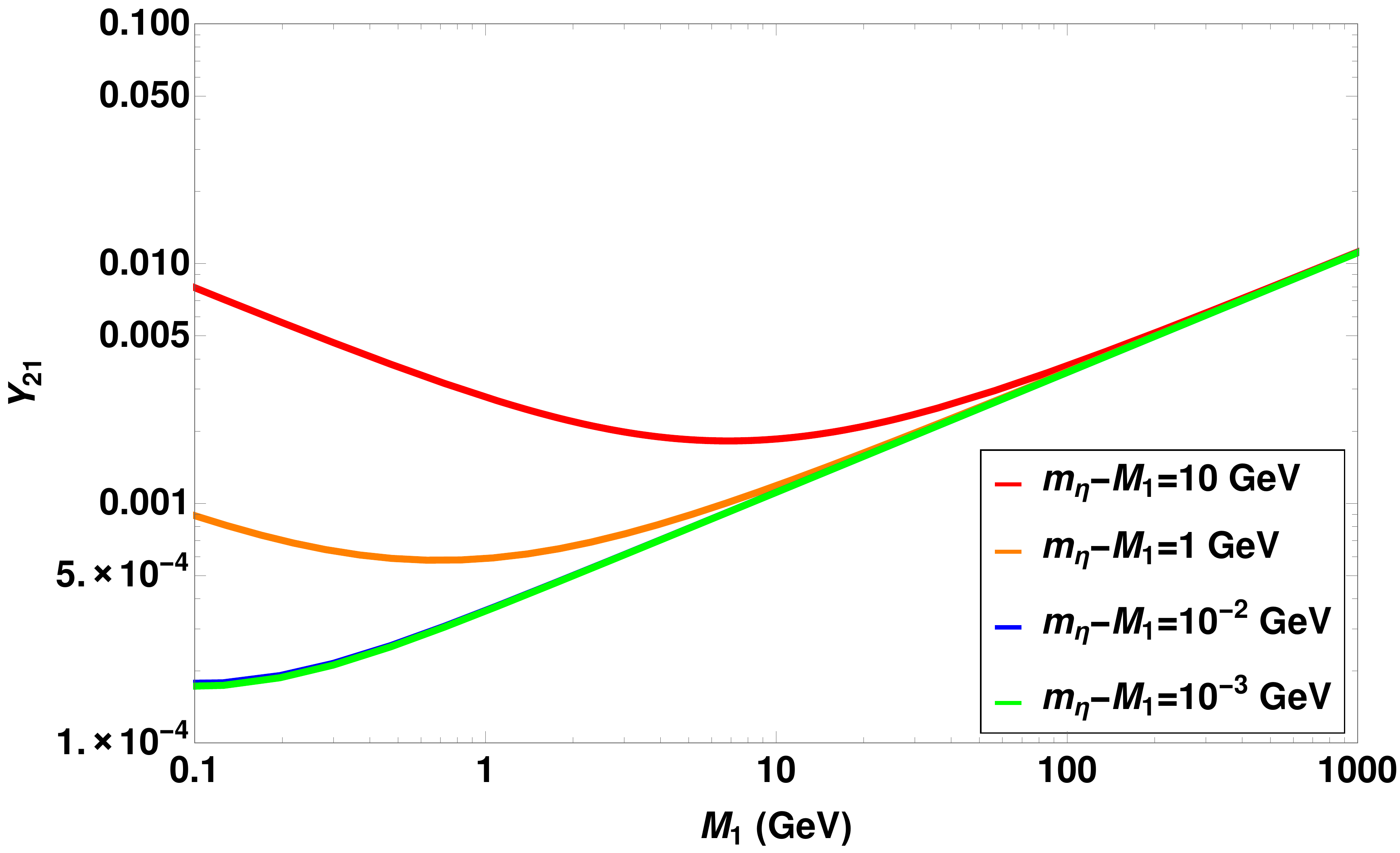} \\
\includegraphics[scale=.22]{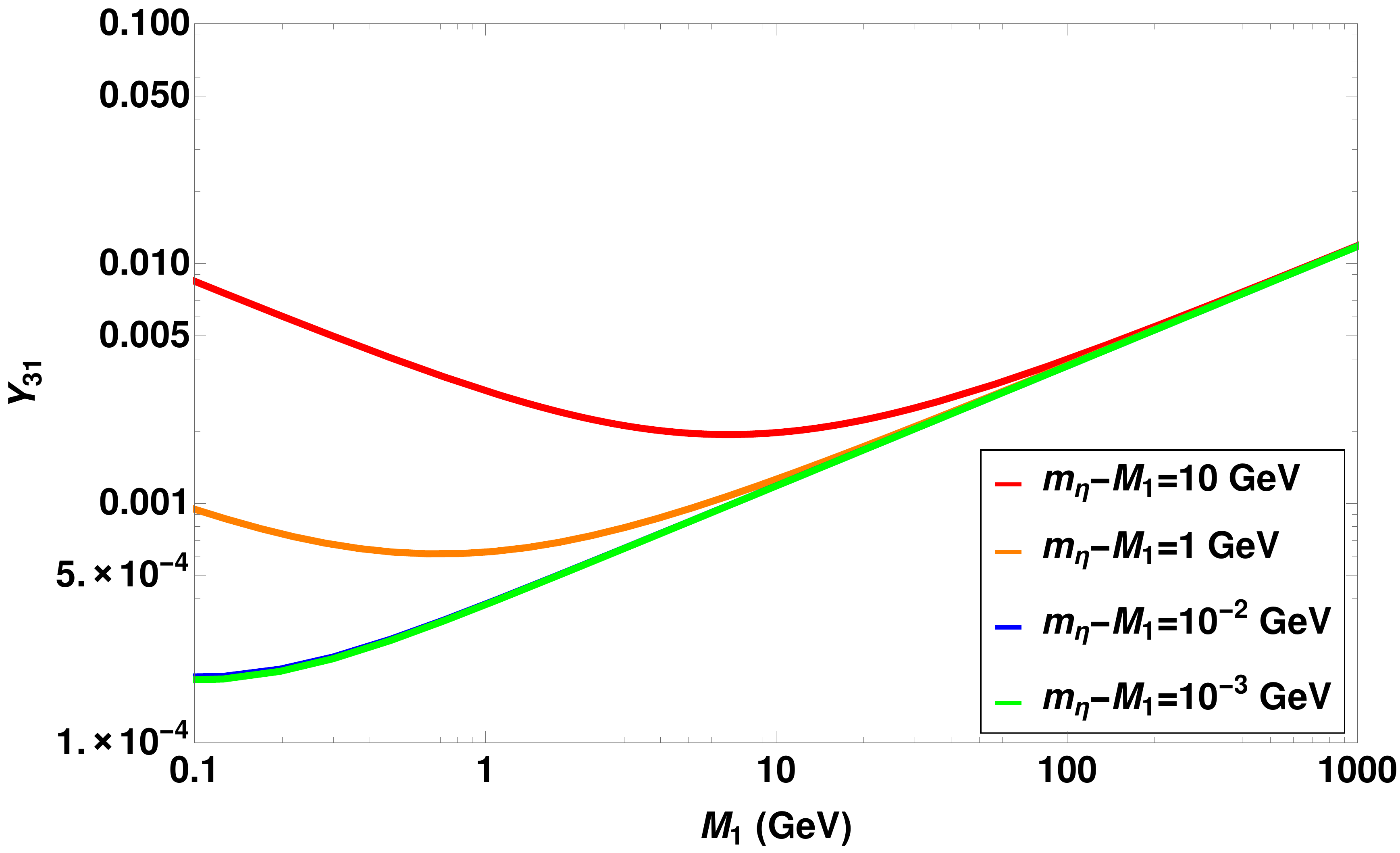}
\caption{Variation of Yukawa coupling with respect to DM mass for IO. The chosen benchmark is $\lambda_5=10^{-6}$.}
\label{yukawaN1IH1}
\end{figure}

As mentioned earlier, we can not have FIMP type Yukawa coupling of DM in IO case due to the structure of Yukawa matrix in terms of light neutrino parameters, for the particular $R$ matrix chosen. We first show the variation of Yukawa couplings of $N_1$ with its mass in figure \ref{yukawaN1IH1}. It can be seen that the couplings can not be made as small as the ones for FIMP dark matter, even though we use the lightest active neutrino mass very small $m_l = 10^{-13}$ eV. We therefore, pursue the WIMP possibility here and show that for small mass splitting between $N_1$ and $\eta$ it is possible to produce the observed relic abundance. The relic abundance of WIMP DM in IO scenario is very similar to the WIMP results obtained in case of NO and hence we do not show the corresponding plot here.

      
\subsection{$N_2$ Leptogenesis with Flavour Effects}    
Finally, we check the role of lepton flavour effects on our results of leptogenesis. In our earlier discussion, we ignored such flavour effects and hence summed over all flavours leading to the CP asymmetry parameter given in \eqref{eq:14}. As pointed out in several earlier works \cite{Abada:2006fw, Abada:2006ea, Nardi:2006fx, Blanchet:2006be}, flavour effects can significantly alter the leptogenesis predictions. For a recent review on flavour effects in leptogenesis, please see \cite{Dev:2017trv}. Here, we adopt the prescription given in \cite{Blanchet:2006be} and calculate the final baryon asymmetry considering a three flavour regime.

\begin{figure}
\includegraphics[scale=.7]{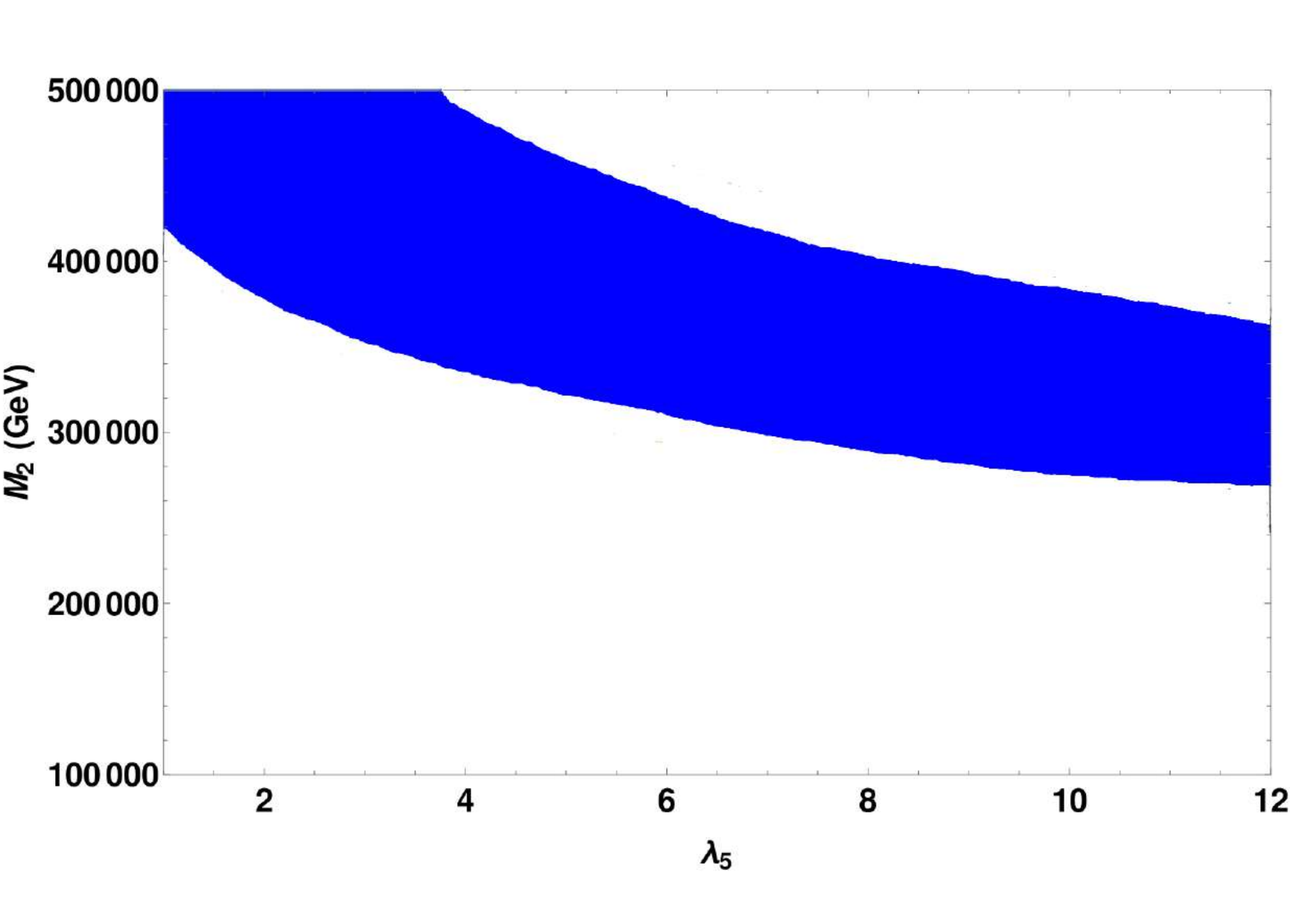}
\caption{Scan plot for flavoured leptogenesis with normal ordering of neutrino mass in $M_{2}-\lambda_{5}$ plane. The benchmark parameters taken for this scan are $m_{1}=10^{-13}$ eV and $M_{3}/M_{2}=10^{5}$.}
\label{scanflavour1}
\end{figure}

\begin{figure}
\includegraphics[scale=.7]{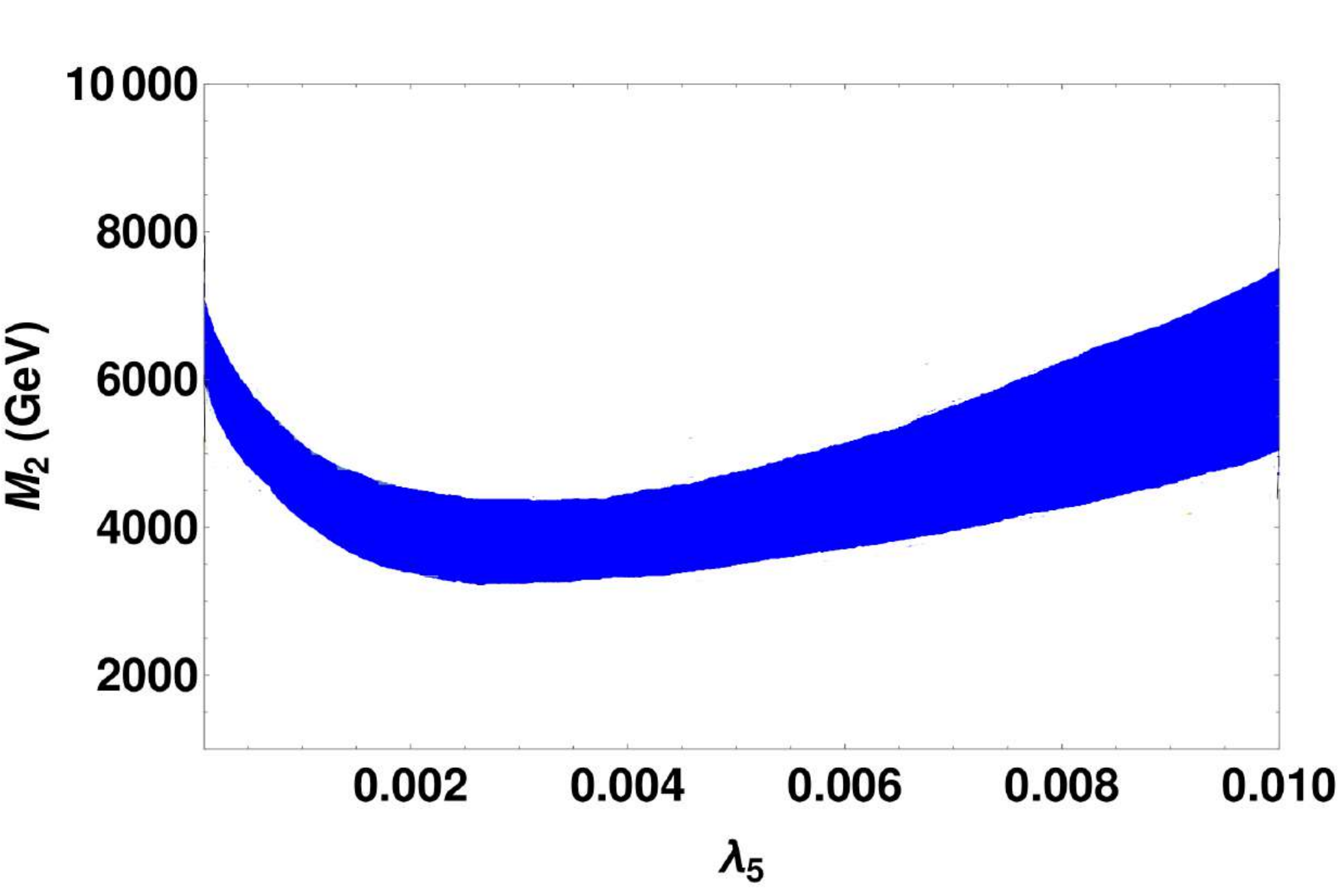}
\caption{Scan plot for flavoured leptogenesis with inverted ordering of neutrino mass in $M_{2}-\lambda_{5}$ plane. The benchmark parameters taken for this scan are $m_{3}=10^{-13}$ eV and $M_{3}/M_{2}=10^{5}$.}
\label{scanflavour}
\end{figure}

The Boltzmann equations for such three flavoured leptogenesis can be written as 
\begin{equation}
\dfrac{dn_{N_{2}}}{dz}=-D_{2}(n_{N_{2}}-n_{N_{2}}^{eq})
\end{equation}
\begin{equation}
\dfrac{dn_{\Delta _{\alpha}}}{dz}=-\epsilon_{2\alpha}D_{2}(n_{N_{2}}-n_{N_{2}}^{eq})-P_{2\alpha}W_{2}^{ID}n_{\Delta_{\alpha}}-P_{2\alpha} \Delta W_{\Delta L=1}n_{\Delta \alpha}-P_{2\alpha} n_{\Delta \alpha} \sum_{\beta=e,\mu,\tau}P_{2\beta}^{0} (\Delta W_{\Delta L=2})_{\alpha \beta}.
\end{equation}
 Here, there are three equations corresponding to $\alpha=e,\mu,\tau$. $n_{N_{2}}$ and $n_{\Delta \alpha}$ are the comoving number densities of $N_{2}$ and the $B/3-L_{\alpha}$ (for each flavor of leptons). $P_{2\alpha}^{0}$ are the projectors defined by 
  \begin{equation}
 P_{2\alpha}=\dfrac{\Gamma_{2\alpha}}{\Gamma_{2}}.
 \end{equation}
 Where $\Gamma_2$ is the total decay width of $N_2$ while $\Gamma_{2\alpha}$ is the corresponding partial decay width to a particular lepton flavour denoted by $\alpha$. The washout terms $\Delta W_{\Delta L=1}, \Delta W_{\Delta L=2}$ are same as the ones discussed earlier. The flavoured CP asymmetry parameter $\epsilon_{2 \alpha}$ is given in \eqref{epsilonflav}.
 We choose the same $R$ matrix as before and find the parameter space for flavoured leptogenesis that can give rise to the correct final baryon asymmetry. The resulting parameter space for NO and IO are shown in figure \ref{scanflavour1} and \ref{scanflavour} respectively. As can be seen from this plot, the scale of leptogenesis can be lower compared to the unflavoured leptogenesis. To be more specific, in case of IO, the leptogenesis scale can be lower by at least an order of magnitude with successful leptogenesis occurring at $M_2$ as low as 300 TeV. On the other hand, in case of IO, the scale of leptogenesis can be as low as a few TeV.

\section{Conclusion}
\label{sec6}
We have studied the possibility of fermion singlet dark matter in the minimal scotogenic model along with explaining the origin of baryon asymmetry of the universe through leptogenesis. The stable nature of the lightest right handed neutrino, being the dark matter candidate, leaves us with the possibility of next to lightest right handed neutrino $N_2$ decay as the source of lepton asymmetry. Compared to the vanilla leptogenesis scenario with $N_1$ decay as main source of lepton asymmetry in minimal scotogenic model, here the scale of leptogenesis gets pushed above, specially in the case of normal ordering. Compared to $M_1 \sim 10$ TeV in $N_1$ decay scenario, here we get similar values $M_1 \sim 20$ TeV for inverted ordering of light neutrinos and $M_1 \sim 10^4$ TeV or normal ordering. We have chosen a particular structure of the complex orthogonal matrix that appears in the Casas-Ibarra parametrisation of Yukawa coupling, the justification for which is given in appendix \ref{appen1}. While the other choices are less efficient in producing the required asymmetry, the chosen structure also explains why it is possible to obtain low scale leptogenesis in inverted ordering scenario while it is not the same with normal ordering. We also point out the importance of new washout terms, which are sub-dominant in vanilla leptogenesis, and show the constraints we get on the lightest neutrino mass $m_{\rm lightest}$ as well as $\lambda_5$, one of the quartic couplings of scotogenic model, from the requirement of producing correct lepton asymmetry.

The correct dark matter relic abundance can be obtained in both the cases either through thermal freeze-out of $N_1$ or freeze-in via decay of $Z_2$ odd scalar doublet $\eta$. In case of thermal freeze-out, the mass splitting between $N_1$ and $\eta$ plays a crucial role in enhancing the coannihilations, bringing the abundance within observed limits. Since WIMP nature of $N_1$ require sizeable Yukawa couplings with $\eta$ to assist coannihilations, and the same Yukawa couplings also play crucial role in washout processes (but not in production) of lepton asymmetry, we remain in the strong washout regime, corresponding to larger values of lightest neutrino mass $m_{1,3}$ discussed earlier. On the other hand, in freeze-in case of NO, the dark matter gets contributions from mother particle $\eta$ while $\eta$ is in thermal equilibrium as well as after $\eta$ freezes out. Here, due to the smallness of Yukawa couplings between $N_1, \eta$ required to implement freeze-in scenario, some of the washout processes remain sub-dominant, lowering the scale of leptogenesis compared to the WIMP scenario. In spite of the scale of leptogenesis being pushed to higher side in NO, there exists rich new physics around the TeV scale in terms of dark matter $N_1$ and the $Z_2$ odd scalar doublet, which can be probed at ongoing experiments. In the end, we check the lepton flavour effects on our leptogenesis results and found that in the three flavour regime, the scale of leptogenesis can be lower by around an order of magnitude compared to the unflavoured leptogenesis. Another interesting prospect of the model is its connection to cosmic inflation. As shown in the recent work \cite{Borah:2018rca}, the $Z_2$ odd scalar doublet $\eta$ can give rise to an inflationary phase of expansion at very early epochs of the universe through its non-minimal coupling to gravity. In the present model also, this remains valid except the fact that there will be additional contribution to reheating as $\eta$ can decay in our present model unlike in \cite{Borah:2018rca} where $\eta$ was considered to be DM and hence stable. We leave exploration of such additional interesting features of our model from both cosmology and particle physics point of view to future works.

\acknowledgments
DM would like to thank Rishav Roshan and Dibyendu Nanda for useful discussions. DB acknowledges the support from IIT Guwahati start-up grant (reference number: xPHYSUGI-ITG01152xxDB001), Early Career Research Award from DST-SERB, Government of India (reference number: ECR/2017/001873) and Associateship Programme of Inter University Centre for Astronomy and Astrophysics (IUCAA), Pune. DB is also grateful to the Mainz Institute for Theoretical Physics (MITP) of the DFG Cluster of Excellence ${\rm PRISMA}^+$ (Project ID 39083149), for its hospitality and its partial support during the completion of this work.
\begin{figure}[h]
\includegraphics[scale=.5]{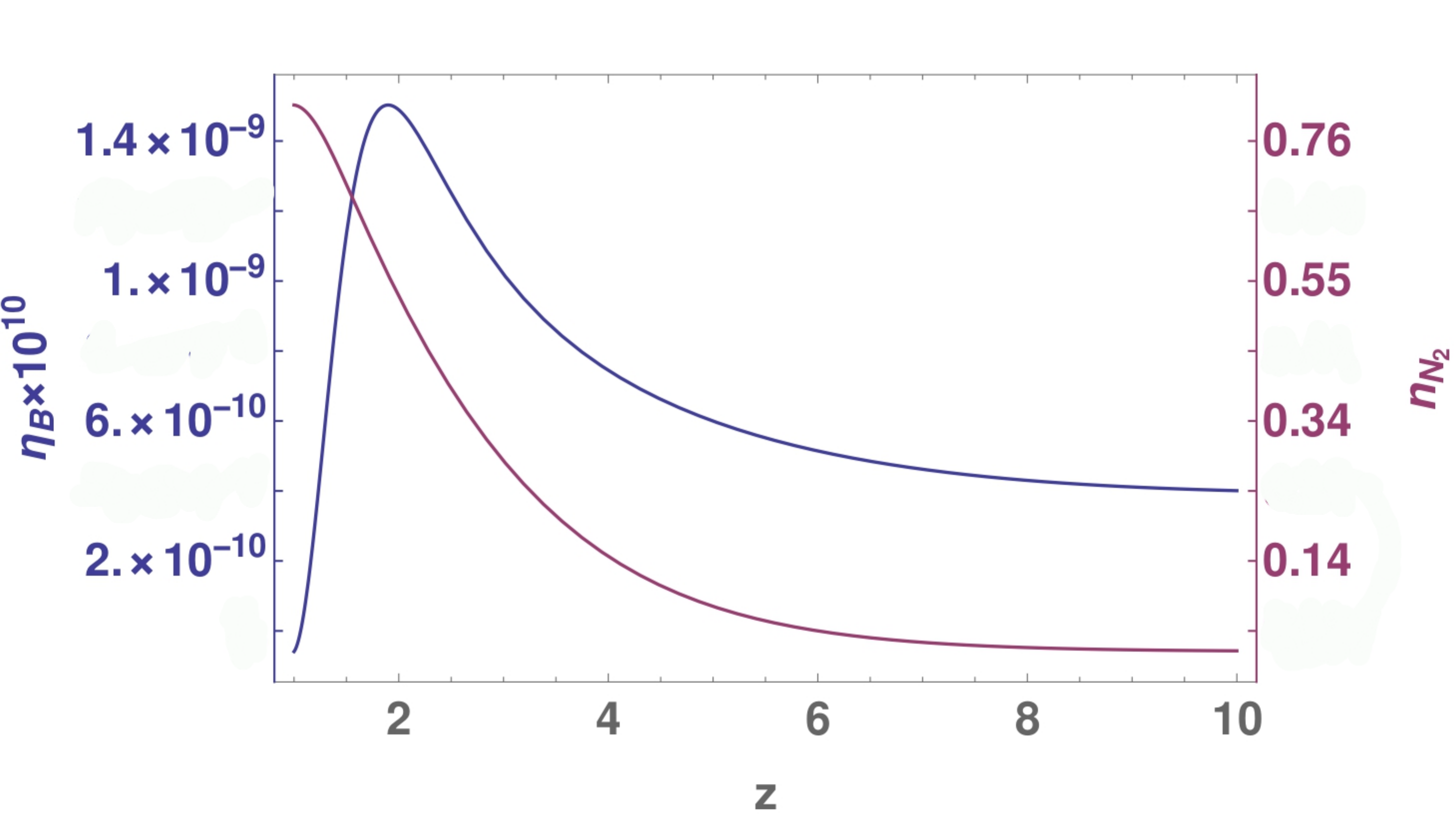}
\caption{Evolution of $n_{N_{2}}$ (Comoving number density of $N_{2}$) and $n_{B-L}$ (Comoving number density of $B-L$) with $z$ for normal ordering and $1-2$ rotation in $R$ matrix. The set of parameters used are $M_{2}=10^{7}$ GeV, $m_{\eta}=450$ GeV, $\lambda_{5}=10^{-4}$ and $M_{1}=10^{3}$ GeV.}
\label{apf1}
\end{figure}
\begin{figure}[h]
\includegraphics[scale=.5]{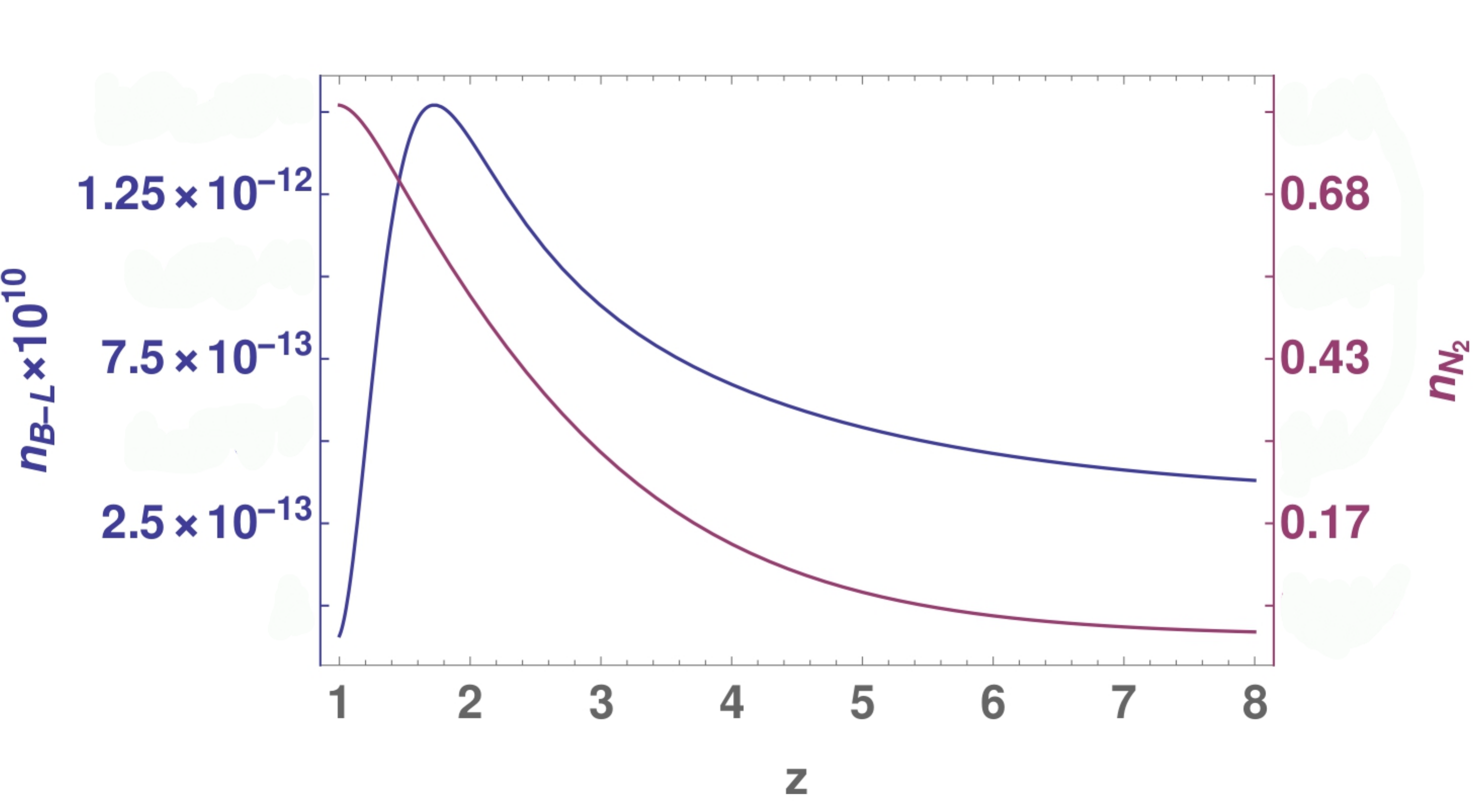}
\caption{Evolution of $n_{N_{2}}$ (Comoving number density of $N_{2}$) and $n_{B-L}$ (Comoving number density of $B-L$) with $z$ for inverted ordering and $1-2$ rotation in $R$ matrix. The set of parameters used are $M_{2}=10^{10}$ GeV, $m_{\eta}=450$ GeV, $\lambda_{5}=0.1$ and $M_{1}=10^{3}$ GeV.}
\label{apf2}
\end{figure}

\appendix
\section{Choice of $R$ matrix and $N_2$ leptogenesis}
\label{appen1}
The choice of complex orthogonal matrix $R$ that appears in the Casas-Ibarra parametrisation of Yukawa couplings \eqref{eq:Yuk}, is crucial for both leptogenesis and dark matter phenomenology. In general, it can be parametrised by three complex parameters. In case of only two right handed neutrinos, the $R$ matrix is a function of only one complex rotation parameter $z=z_R + i z_I, z_R \in [0, 2\pi], z_I \in \mathbb{R}$ \cite{Ibarra:2003up}. This does not leave much freedom in choosing $R$ and gives rise to a lower bound on the scale of leptogenesis very similar to the Davidson-Ibarra bound $M_1 > 10^9$ GeV \cite{Davidson:2002qv} even in scotogenic model with two right handed neutrinos \cite{Hugle:2018qbw}. However, in our case, although leptogenesis is due to $N_2$ decay, we still have more freedom in choosing $R$ compared to the two right handed neutrino scenario. As discussed in the main text, our choice of $R$ matrix is 
\begin{equation}
R=\begin{pmatrix}
1&0&0\\
0& \cos{z} & \sin{z}\\
0& -\sin{z} & \cos{z}
\end{pmatrix}
\end{equation}
Recalling the relation between Yukawa and $R$ \eqref{eq:Yuk} that is, $ Y \ = \ U D_\nu^{1/2} R^{\dagger} \Lambda^{1/2}$ and the product of Yukawas relevant for CP asymmetry $(Y^{\dagger}Y)_{ij}=\sqrt{\Lambda_{i}\Lambda_{j}}(RD_{\nu}R^{\dagger})_{ij}$, we calculate, for the above choice of $R$ matrix, the following quantity
\begin{equation}
RD_{\nu}R^{\dagger}=\begin{pmatrix}
m_1&0&0\\
0&m_2 \cos{z} (\cos{z})^*+m_3 \sin{z} (\sin{z})^* & -m_2 (\sin{z})^* \cos{z} + m_3 (\cos{z})^* \sin{z} \\
0& -m_2 (\cos{z})^* \sin{z} + m_3 \cos{z} (\sin{z})^* & -m_2 (\sin{z})^* \sin{z} + m_3 \cos{z} (\cos{z})^*
\end{pmatrix}
\label{eqap1}
\end{equation}
This clearly gives a non-zero complex entry in $(Y^{\dagger}Y)_{23}$ which will contribute to net CP asymmetry $\epsilon_2$ in accordance with equation \eqref{eq:14}. This choice of $R$ also explains the reason behind the difference in the scale of leptogenesis we obtained for NO and IO. As can be seen from the CP asymmetry parameter $\epsilon_2$ given in \eqref{eq:14}, it is also inversely proportional to $(Y^{\dagger}Y)_{22}$. Therefore, for maximum CP asymmetry $(Y^{\dagger}Y)_{22}$ should be smaller and imaginary part of $((Y^{\dagger}Y)_{23})^2$ should be larger. As can be seen from the matrix given in equation \eqref{eqap1}, the $(22)$ element can be made very small for IO by choosing $z$ in such a way that makes $\cos{z}$ small. The term containing $\sin{z}$ can be small by choosing $m_3$ arbitrarily small. Since we are not making $\sin{z}$ arbitrarily small, we can still have a larger $(23)$ term of the matrix \eqref{eqap1} to enhance the CP asymmetry parameter $\epsilon_2$. However, in case of NO, we can not choose either $m_2$ or $m_3$ to be small and hence it is not possible to get a hierarchy between $(23)$ and $(22)$ terms of the matrix \eqref{eqap1}. Therefore, the only way that can increase the CP asymmetry parameter is by pushing the scale $M_2$ up. This results in higher scale of leptogenesis in NO compared to that in IO.

If we had taken a different choice of $R$, with the rotation parameters either in $1-3$ plane or $1-2$ plane, we will get
\begin{equation}
RD_{\nu}R^{\dagger}=\begin{pmatrix}
m_1 \cos{z} (\cos{z})^* + m_3 \sin{z} (\sin{z})^* & 0 & -m_{1} \cos{z} (\sin{z})^* + m_3 (\cos{z})^* \sin{z} \\
0& m_{2} & 0\\
-m_1 (\cos{z})^* \sin{z} + m_3 \cos{z} (\sin{z})^* & 0 & m_1 (\sin{z})^* \sin{z} + m_3 \cos{z} (\cos{z})^*
\end{pmatrix}
\label{ap3}
\end{equation}
and
\begin{equation}
RD_{\nu}R^{\dagger}=\begin{pmatrix}
m_1\cos{z} (\cos{z)})^*+m_2\sin{z} (\sin{z})^* & -m_{1}\cos{z} (\sin{z})^*+m_3 (\cos{z})^* \sin{z} & 0\\
-m_1(\cos{z})^* \sin{z} +m_2\cos{z} (\sin{z})^* & m_1\sin{z} (\sin{z})^*+m_3\cos{z} (\cos{z)})^* & 0\\
0&0&m_3
\end{pmatrix}
\label{ap4}
\end{equation}
respectively.
Clearly, the rotation only in $1-3$ plane can not give rise to non-vanishing CP asymmetry in our case, as both $(Y^{\dagger}Y)_{23}$ and $(Y^{\dagger}Y)_{21}$ terms appearing in CP asymmetry formula \eqref{eq:14} are vanishing as seen from \eqref{ap3}. A rotation in $1-2$ plane can however, give rise to a net CP asymmetry, as seen from \eqref{ap4}. We now try to estimate the strength of the resulting lepton asymmetry from such a choice of $R$. Let us choose the $R$ matrix to be 
\begin{equation}
R=\begin{pmatrix}
\cos{z} & \sin{z} & 0\\
-\sin{z} & \cos{z} & 0\\
0&0&1
\end{pmatrix}
\end{equation}
with $z=0.82+1.42i$ for NO and $z=0.48-0.58i$ for IO. We then solve the coupled Boltzmann equations to find the evolution of $N_2$ number density and $B-L$ asymmetry for both NO and IO. The resulting plots are shown in figure \ref{apf1} and \ref{apf2} respectively. As can be seen from these two plots, the net lepton asymmetry generated for such a choice of $R$ matrix remain several order of magnitudes smaller than the required one \footnote{Here, for simplicity, we have not considered $\Delta L=1$ washout processes, including which, will lower the asymmetries further.}. Therefore, it justifies the use of $2-3$ rotation in $R$ matrix as was done in the main text. We also check that, it still remains suppressed even if we push the scale of leptogenesis higher say $M_2 \sim 10^{14}$ GeV. Apart from the $R$ matrix, another factor which affects the resulting asymmetry is the loop function $F(r_{ji}, \eta_i)$ in CP asymmetry formula \eqref{eq:14}. For $1-2$ rotation, it is effectively the contribution from $N_1$ in loop which is contributing the net CP asymmetry from $N_2$ decay. Since $N_1$ is lighter than $N_2$ we have $r_{ji} \equiv r_{12} <1$ and the loop factor $F(r_{12}, \eta_2)$ gets suppressed in this regime. On the other hand for $2-3$ rotation the loop factor $F(r_{32}, \eta_2)$ can be large as we are in the regime $r_{ji} \equiv r_{32} >1$.

Now, coming to the implications for dark matter sector, let us consider the $R$ matrix to be a multiplication of two different rotation matrices $R=R_{23}R_{13}$ given by 
\begin{equation}
R=\begin{pmatrix}
\cos{z'} & 0 & \sin{z'}&\\
-\sin{z} \sin{z'} & \cos{z} & \sin{z} \cos{z'}\\
-\cos{z} \sin{z'} & -\sin{z} & \cos{z} \cos{z'}
\end{pmatrix}
\end{equation}
This choice of $R$ matrix will give us the following Yukawa couplings for $N_{1}$ to the three lepton generations 
\begin{equation}
Y_{i1}=\begin{pmatrix}
\sqrt{m_1}\sqrt{\Lambda_{1}}(\cos{z'})^*U_{11}+\sqrt{m_{3}}\sqrt{\Lambda_{1}}(\sin{z'} )^*U_{13}\\
\sqrt{m_1}\sqrt{\Lambda_{1}}(\cos{z'})^*U_{21}+\sqrt{m_{3}}\sqrt{\Lambda_{1}}(\sin{z'} )^*U_{23}\\
\sqrt{m_1}\sqrt{\Lambda_{1}}(\cos{z'})^*U_{31}+\sqrt{m_{3}}\sqrt{\Lambda_{1}}(\sin{z'} )^*U_{33}
\end{pmatrix}
\end{equation}
where $U_{ij}$ are the PMNS matrix elements. If we set $z'=0$, we recover the first column of Yukawa matrix given in equation \eqref{yukawamatrix1}. In that case, as we mentioned earlier, if we have normal ordering of light neutrino masses, we can have small Yukawa couplings of $N_1$ by choosing small $m_1$. Or else, we can choose sizeable Yukawa by choosing large values of $m_1$. These two scenarios can lead to thermal and non-thermal dark matter possibilities respectively. Now, for inverted ordering, we can not have arbitrarily small Yukawa in the $z'=0$ limit which we discussed in the main text above. Since for inverted ordering $m_3$ can be arbitrarily small, we can choose $z'$ in such a way that $\cos{z'}$ is very small. This can in principle give rise to tiny Yukawa couplings of $N_1$ in inverted ordering case, realising the non-thermal dark matter scenario. Since $13$ rotation parameter $z'$ does not produce non-vanishing CP asymmetry as mentioned earlier, we did not discuss it in this work.

\providecommand{\href}[2]{#2}\begingroup\raggedright\endgroup

\end{document}